\documentclass[fleqn,usenatbib]{mnras}
\usepackage[T1]{fontenc}
\usepackage[english]{babel}
\usepackage{txfonts}
\usepackage[abs]{overpic}
\usepackage{pict2e}

\usepackage{graphicx}	

\def\l{\ensuremath{\lambda}}

\newcommand{\tim}[1]{\ensuremath{\times 10^{#1}}}

\def\apec{{\sc apec}}

\def\xspec{{\sc xspec}}

\newcommand{\ergs}{erg\,cm$^{-2}$s$^{-1}$}
\newcommand{\ergpers}{erg\,s$^{-1}$}

\def\Ebv{{\ensuremath{E_{B-V}}}}

\def\HST{\emph{HST}}
\def\GALEX{\emph{GALEX}}

\def\WISE{\emph{WISE}}
\def\NEOWISE{\emph{NEOWISE}}

\def\Rwd{\ensuremath{R_{\rm WD}}}
\def\Mwd{\ensuremath{M_{\rm WD}}}

\def\cd{cycle d$^{-1}$}

\newcommand{\vsini}{$V_{\rm rot}\sin i$}
\newcommand{\Msun}{M$_{\sun}$}
\newcommand{\Rsun}{R$_{\sun}$}
\newcommand{\Teff}{$T_{\rm eff}$}
\newcommand{\Porb}{$P_{\rm orb}$}

\newcommand{\kms}{km\,s$^{-1}$}
\newcommand{\myemail}{vitaly@neustroev.net}
\newcommand{\iraf}  {{\sc iraf}}
\newcommand{\PyA}  {{\sc PYASTRONOMY}}

\newcommand{\ESOREFLEX} {{\sc ESO reflex}}
\newcommand{\MOLECFIT} {{\sc molecfit}}
\newcommand{\pessto} {{\sc pessto}}

\newcommand {\bc}{\begin {center}}
\newcommand {\ec}{\end {center}}
\newcommand {\be}{\begin {equation}}
\newcommand {\ee}{\end {equation}}
\newcommand {\beq}{\begin {eqnarray}}
\newcommand {\eeq}{\end {eqnarray}}

\newcommand{\Halpha} {H$\alpha$}
\newcommand{\Hbeta}  {H$\beta$}
\newcommand{\Hgamma} {H$\gamma$}
\newcommand{\Hdelta} {H$\delta$}
\newcommand{\Hepsilon} {H$\varepsilon$}
\newcommand{\Hzeta} {H$\zeta$}
\newcommand{\Paalpha} {Pa$\alpha$}

\newcommand{\Pagamma} {Pa$\gamma$}

\newcommand{\Brgamma} {Br$\gamma$}

\newcommand{\CaII} {Ca\,{\sc ii}}

\newcommand{\CI}  {C\,{\sc i}}

\newcommand{\FeI} {Fe\,{\sc i}}

\newcommand{\HeI}  {He\,{\sc i}}
\newcommand{\HeII} {He\,{\sc ii}}

\newcommand{\MgII} {Mg\,{\sc ii}}
\newcommand{\KI}   {K\,{\sc i}}
\newcommand{\NaI}  {Na\,{\sc i}}

\newcommand{\OI}   {O\,{\sc i}}

\newcommand{\SiI} {Si\,{\sc i}}

\title[Spectroscopy and photometry of BW Scl]
{
A brown dwarf donor and an optically thin accretion disc with a complex stream impact region
in the period-bouncer candidate BW Sculptoris
}

\author[V.~V.~Neustroev \& I.~M{\"a}ntynen]{Vitaly~V.~Neustroev$^{1}$\thanks{E-mail: \myemail},
Iikka~M{\"a}ntynen$^{1,2}$
\\
$^{1}$Space Physics and Astronomy research unit, PO Box 3000, FI-90014 University of Oulu, Finland\\
$^{2}$Department of Physics and Astronomy, University of Turku, FI-20014 Turku, Finland\\
}

\date{Accepted 2023 June 5. Received 2023 June 4; in original form 2022 August 23}

\pubyear{2023}

\begin{document}
\label{firstpage}
\pagerange{\pageref{firstpage}--\pageref{lastpage}}
\maketitle

\begin{abstract}
We present an analysis of multi-epoch spectroscopic and photometric observations of the WZ~Sge-type dwarf nova BW~Scl,
a period-bouncer candidate. We detected  multiple irradiation-induced emission lines from the donor star allowing
the radial velocity variations to be measured with high accuracy. Also, using the absorption lines Mg\,{\sc ii} 4481~\AA\
and \CaII~K originated in the photosphere of the accreting white dwarf (WD), we measured the radial velocity semi-amplitude
of the WD and its gravitational redshift. We find that the WD has a mass of 0.85$\pm$0.04~M$_\odot$, while the donor is a low-mass
object with a mass of 0.051$\pm$0.006~M$_\odot$, well below the hydrogen-burning limit. Using NIR data, we put an upper
limit on the effective temperature of the donor to be $\lesssim$1600~K, corresponding to a brown dwarf of T spectral type.
The optically thin accretion disc in BW~Scl has a very low luminosity $\lesssim$4\tim{30} \ergpers\ which corresponds to
a very low-mass accretion rate of $\lesssim$ 7\tim{-13} \Msun\ yr$^{-1}$. The outer parts of the disc have a low density
allowing the stream to flow down to the inner disc regions. The brightest part of the hotspot is located close to the
circularization radius of the disc. The hotspot is optically thick and has a complex elongated structure.
Based on the measured system parameters, we discuss the evolutionary status of the system.
\end{abstract}


\section{Introduction}

Cataclysmic variables (CVs) are interacting binaries in which the primary, a white dwarf (WD) accretes matter from the secondary,
a late-type donor star \citep{Warner}. The standard evolution of CVs is driven by angular momentum loss, which results in CVs
evolving from longer orbital periods to shorter ones before reaching a minimum period of $P_{\rm min}$$\simeq$70--80 min
\citep{Kolb93,Howell97,Howell01,KniggeCVevol,Belloni20}. Approaching $P_{\rm min}$, the donor star evolves
into a sub-stellar, brown dwarf-like object, at which point the orbital period $P_{\rm orb}$ of the binary starts increasing
again. CVs that have passed beyond $P_{\rm min}$ and are bouncing back towards longer periods are called `period bouncers'.

Population models predict a vast number of highly evolved CVs with sub-stellar donors (up to 80\% of the present day
Galactic CV population -- \citealt{Belloni20}; see also \citealt{Kolb93,GoliaschNelson,Schreiber16}), whereas only about two
dozens of more or less robust candidates for
these predicted objects have been identified until now \citep[for published compilations see][]{PattersonDist,Kimura18}.
Moreover, direct (spectral) evidence for brown dwarf donors exists only for very few CVs (SDSS J143317.78+101123.3;
\citealt{HernandezSantisteban16}; WZ~Sge; \citealt{HarrisonWZ,HarrisonWZ2}; SDSS J123813.73-033933.0;
\citealt{Pala-SDSS1238}). The recognition of the sub-stellar nature of the donors in other period-bounce candidates
is based on indirect methods such as radial velocity measurements of the WDs, the use of a superhump period--mass ratio
relation and analysis of the spectral energy distribution (SED).

An accurate characterization of donor stars in old CVs is a challenging task because the donor photometric properties along
the CV evolution sequence change dramatically. In period bouncers, a brown dwarf donor of very late spectral type (L/T) is
expected \citep{KniggeCVevol}. Such a donor star is so dim that even a weak contribution from the accretion disc and the
accretion-heated WD can easily outshine it. \citet{Littlefair03} suggest that the chances of detecting the brown dwarf
donors are increased in nearby CVs with little or no ongoing accretion.

\begin{table*}
\begin{center}
\caption{Log of spectroscopic observations of BW~Scl.}
\begin{tabular}{clccccc}
\hline\hline
 HJD-Start & Telescope /     & Sp. res &  \l~range  & Exp.time & Number   &  Duration   \\
 2450000+  & Instrument      &   R   &     (\AA)    &  (s)     & of exps. & (h / \Porb) \\
\hline
 2227.4977 & VLT / UVES      & 44000 &  4020--5245  &  300     &  60      & 5.53 / 4.24 \\
 2228.5051 & VLT / UVES      & 44000 &  3300--4520  &  300     &  15      & 1.30 / 1.00 \\
 2228.5057 & VLT / UVES      & 44000 &  4625--5600  &  300     &  15      & 1.30 / 1.00 \\
 2228.5057 & VLT / UVES      & 46000 &  5680--6650  &  300     &  15      & 1.30 / 1.00 \\
 2228.5643 & VLT / UVES      & 44000 &  4020--5245  &  300     &  30      & 2.68 / 2.05 \\
 2494.7781 & VLT / UVES      & 44000 &  4020--5245  &  300     &  34      & 4.15 / 2.41 \\
 2495.8042 & VLT / UVES      & 44000 &  4020--5245  &  300     &   8      & 0.68 / 0.52 \\


 5482.4935 & VLT / X-shooter &  4300 &  3000-- 5560 &   60     &  239     & 5.54 / 4.25 \\
 5482.4935 & VLT / X-shooter &  7400 &  5560--10200 &   60     &  239     & 5.54 / 4.25 \\
 5482.4935 & VLT / X-shooter &  5400 &  9950--24780 &   60     &  239     & 5.54 / 4.25 \\


 8030.5847 & NTT / SOFI      &  1000 &  9500--16400 &  180     &  26      & 1.32 / 1.01 \\
 8385.6834 & NTT / EFOSC2    &   800 &  3300--5200  &  300     &   1      & 0.08 / 0.06 \\
 8385.6868 & NTT / EFOSC2    &  1000 &  4700--6770  &  180     &  22      & 1.30 / 1.00 \\
 8385.7422 & NTT / EFOSC2    &   600 &  4045--7445  &  300     &   1      & 0.08 / 0.06 \\
\hline

\end{tabular}
\label{Tab:SpecObs}
\end{center}
\end{table*}

BW~Sculptoris, a $\sim$16.5 mag blue star, is one of the brightest and closest CVs \citep[$d$=93.3$\pm$0.4 pc --][]{GaiaEDR3,GaiaDist}
included to the period-bouncer scorecard by \citet{PattersonDist}. It was discovered as a soft X-ray source in the ROSAT all-sky
survey \citep[RX~J2353.0-3852;][]{Abbott97} and independently found in Hamburg/ESO survey for bright quasars
\citep[HE~2350-3908;][]{Augusteijn97}. These two studies established that BW~Scl is a CV with the very short orbital period
of $\sim$78.2 min, the shortest known to that date. Photometric observations performed by \citet{Uthas12} from 1999 to 2009 resulted
in an accurate measurement of the orbital period $P_{\rm orb} = 78.22639\pm0.00003$ min. In addition to double-peaked emission
lines, the spectrum of BW~Scl also shows very broad Lyman and Balmer absorption lines from the underlying accreting WD of
modest temperature ($\sim$15000~K -- \citealt{Gaensicke05,Pala22}). Such
a spectrum was found to be very similar to WZ~Sge, which allowed \citet{Augusteijn97} to argue that the source is a dwarf nova
with a long recurrence time and a very low mass-transfer rate. This motivated \citet{Mennickent04} to conduct a search for a brown
dwarf in BW~Scl, but near-infrared (NIR) spectroscopy has failed to reveal the donor in this system.

In October 2011, BW~Scl experienced a superoutburst of an amplitude of $\sim$7.5 mag \citep{Kato4}, confirming thus a WZ~Sge-type
classification \citep[for a review of the WZ~Sge-type dwarf novae, see][]{KatoWZ}. Based on their method of
estimating binary’s mass ratios by using the period of superhumps, \citet{KatoOsaki13} reported a mass ratio of BW~Scl to be
$q \equiv M_2 / \Mwd = 0.067\pm0.006$, where $M_2$ and \Mwd\ are masses of the donor and the WD, respectively. An even lower mass
ratio ($q$ = 0.062) can be obtained from another empirical relation between $q$ and the orbital and superhump
periods \citep{McAllister19}. Together with a mass estimate for the WD of 1.01 \Msun\ derived by \citet{Pala22} from the
re-analysis of the \emph{Hubble Space Telescope} (\HST) data \citep{Gaensicke05}, this yields the donor mass to be
0.062--0.068~\Msun. This mass range does not allow for a certain confirmation of the donor's brown dwarf status because
it is only slightly lower than the hydrogen-burning mass limit \citep{H-limit-1,H-limit-2}. Still, these estimates are based
on indirect estimates from relations which are not well calibrated at very low-mass ratios.

This motivated us to perform a detailed spectroscopic study of BW~Scl aiming in a more reliable determination of the
fundamental system parameters. As a result, we confirm the sub-stellar nature of the donor star and argue that
BW~Scl has already passed the minimum period and is now evolving back towards longer periods.

\section{The data}

The quantitative analysis presented in this work is based primarily on the spectra obtained with different telescopes and
instruments of the European Southern Observatory (ESO). The full journal of spectroscopic observations is presented in
\autoref{Tab:SpecObs}. These data combine new and archival material. The new observations are photometry and spectroscopy
taken with the New Technology Telescope (NTT) at La Silla Observatory (Chile). We complement our observations with archival
spectroscopic data which were obtained with the Very Large Telescope (VLT) at the Paranal Observatory in Chile, using the
medium resolution spectrograph X-shooter and the Ultraviolet and Visual \'{E}chelle Spectrograph (UVES).
See Section~\ref{Sec:SpecData} for more detail regarding these data.

Our data sets also comprise new observation of BW~Scl with the Neil Gehrels Swift Observatory (Section~\ref{Sec:XrayData}).
Additionally, we employed high-level science products derived from some other archives, which we describe in due course.

\subsection{Optical and NIR photometry and spectroscopy}
\label{Sec:SpecData}

We used the NTT telescope to perform multicolour optical and NIR photometry and spectroscopy of BW~Scl (programs 100.D-0932 and
101.D-0806; PI: V. Neustroev). The NIR observations were made on 2017 October 4 with the infrared spectrograph and imaging camera
SOFI 
\citep{MoorwoodSOFI} mounted at the Nasmyth A focus
of the NTT. The NIR images were obtained through the JHK$_{\rm s}$ filters, while the spectra were taken with
the blue grism covering the wavelength range 9500--16400~\AA. The data were taken under clear atmospheric conditions with seeing
$\sim$1 arcsec. In order to increase the resolving power to about R$\sim$1000, we used a narrow slit of 0.6 arcsec.
Spectra of the telluric standard Hip 117513 were taken before and after the observations of the target for flux calibrations.
A comparison spectrum of a xenon lamp was used for wavelength calibration. The SOFI data were reduced using the \pessto\ pipeline
version 2.4.1 \citep{PESSTO}.

The optical data were acquired with the ESO Faint Object Spectrograph and Camera
EFOSC2 
\citep{BuzzoniEFOSC} mounted at the
Nasmyth B focus of the NTT. The images were captured on 2017 October 5 through the BVRiz filters. The spectroscopic observations
were performed on 2018 September 24 under perfect weather conditions with seeing 0.6--0.7 arcsec, allowing us to use a narrow
slit of 0.7 arcsec. The time-resolved observations were taken with grism \#18 in the wavelength range of 4700--6770 {\AA} with
a spectral resolution of 5.7~\AA. A total of 22 spectra with 180 s individual exposures were obtained, covering one orbital
period of the system. In order to maximize the wavelength coverage, we also took two additional spectra using grisms \#4 and \#7
with the exposure times of 300~s, providing an overall wavelength coverage of 3270--7520~\AA.
For flux calibrations, spectra of the standard spectrophotometric star LTT 7987 were taken. For wavelength calibrations, He-Ar
lamp exposures were obtained. The EFOSC2 data were reduced using \iraf\ package in a standard way.

The optical magnitudes of BW~Scl in different filters were measured using four comparison stars in the field. Their magnitudes
were obtained from the VST ATLAS survey\footnote{\url{https://astro.dur.ac.uk/Cosmology/vstatlas/}} \citep{VST}, DR4 data release.
The NIR magnitudes were calibrated using the same comparison stars as were chosen for the optical photometry. Their magnitudes
and errors were taken from the 2MASS All-Sky Catalog of Point Sources \citep{2MASS}. The derived magnitudes of BW~Scl are shown
in Table~\ref{Tab:Magnitudes}. In the following, we call the data which were obtained with the NTT as the NTT data set, or the
EFOSC2 or SOFI spectra if necessary.

BW~Scl was observed twice with VLT/UVES \citep{UVES} in 2001 and 2002.
The observations obtained on 2001 November 13--15 under program 068.D-0153 were performed using different configurations
allowing to cover the wavelength range from 3300 to 6650~\AA. All segments of the spectrum cover at least 1 orbital period
of the system, while blue-arm spectra (4020--5245~\AA) cover $\sim$6.3~\Porb. The data taken on 2002 August 08--09 under program
069.D--0391 consist of only blue-arm spectra covering $\sim$2.9~\Porb. All the UVES observations were acquired using a 0.9 arcsec
slit yielding a resolving power of approximately 44000. An exposure time was 300~s.
These data are referred to as the UVES data set, or the UVES-1 or UVES-2 spectra in case it is
necessary to specify which observations (either obtained in 2001 or 2002) are used.

On 2010 October 12--13, BW~Scl was also observed with the medium resolution spectrograph X-shooter (program ID 086.D-0775).
This instrument is comprised of three detectors \citep{Xshooter}: the UVB arm, covering 3000--5500 \AA; the VIS arm, covering
5500--10000 \AA; the NIR arm, covering 10000--25000 \AA. The observations used slit widths of 1.0 arcsec, 0.9 arcsec and
0.9 arcsec for the UVB, VIS, and NIR arms, respectively, and 2$\times$2 binning in the UVB and VIS arms. This resulted in
a resolving power of 4300, 7400, and 5400 in the UVB, VIS, and NIR arms, respectively. A total of 239 spectra with 60 s
individual exposures were obtained, covering $\sim$4.2 orbital periods of the system. The observations were performed in
STARE mode. We call this data set the X-shooter data, or the UVB, VIS, or NIR spectra.

The raw UVES and X-shooter data together with calibration files were retrieved from the ESO Science Archive
Facility and reduced using the standard pipelines (UVES Workflow For Point Source Echelle
Data version 6.1.6, and X-shooter Workflow for Physical Mode Date Reduction version 3.5.0) within \ESOREFLEX\ version 2.11.5
\citep{Freudling13}. The standard recipes were used to optimally extract and wavelength and flux calibrate each spectrum.
Individual X-shooter VIS and NIR spectra were separately corrected for telluric absorption using \MOLECFIT\
\citep{Molecfit1,Molecfit2}. Finally, the wavelength scale of all the spectra of each data set was also barycentrically
corrected.

In addition to individual spectra, we also produced flux-calibrated average spectra, which are a combination of all spectra
from the corresponding data sets, uncorrected for orbital motion.\footnote{The average X-shooter
spectra were obtained by means of coadding science frames acquired within the same Observation Block (OB) \emph{before} reducing them
(the rule xsh\_wkf\_starestack.oca in \ESOREFLEX). This approach was especially useful for the NIR arm, because the individual
NIR spectra have a very low signal-to-noise ratio (SNR) $\sim$1 in continuum.}
Because the obtained mean spectra can suffer from imperfect flux calibration,\footnote
{A flux calibration method based on observations of spectrophotometric standards is subject of uncertainties due to slit losses
and differences in seeing and weather conditions between observations of the target and standards.}
we scaled them to match the fluxes derived from either quasi-simultaneous multicolour photometry (the NTT data) or acquisition
images (the X-shooter data). Unfortunately, no acquisition images are available for the UVES data. Nevertheless, we show
in Section~\ref{Sec:LongTermPhot} that during the UVES and X-shooter observations BW~Scl was at nearly the same flux level.
Since most of our mean spectra span a broad wavelength range, which is covered by several photometric
passbands, we also attempted to correct the spectra for wavelength-dependent effects such as atmospheric transmission.
For this, we followed an approach similar to that used e.g. by \citet{FluxCalib}.
The mean spectra were convolved with the filter bandpasses to determine the ratios of the photometric fluxes in each filter to the
fluxes measured in the spectra. The ratios were then fitted by a low-order polynomial to establish a correction function.
As a result, we have reached the flux calibration to be accurate within a few per cent in the whole wavelength range.\footnote{
Note that although the mean UVES-2 spectrum has a relatively high SNR of $\sim$100 in continuum, it clearly
shows the \'{E}chelle order pattern (\autoref{Fig:UVES}). \citet{Korn07} pointed out that the blaze removal and order merging as
implemented in the UVES pipeline leave significant residuals which may affect the analysis of WD lines.}
Finally, BW~Scl is a nearby source and its interstellar extinction of \Ebv = 0.002$\pm$0.015 \citep{Dereddening} is negligible
for the analysis presented here \citep[see][]{Pala17}, thus no dereddening correction was applied.

\begin{table*}
\caption{The magnitudes of BW~Scl derived from the NTT and {\it Swift}-UVOT observations. They are given in the Vega system 
except for the $i$ and $z$ magnitudes which are in the AB system.}
\begin{center}
\begin{tabular}{cccccccccccc}
\hline
 $uvw2$   &  $uvm2$  & $uvw1$   &   $u$    &   $B$    &  $V$     &   $R_c$  &   $i$    &   $z$    &   $J$    &   $H$    &   $Ks$   \\
\hline
 14.97(2) & 15.02(4) & 15.22(4) & 15.55(4) & 16.60(3) & 16.61(3) & 16.37(1) & 16.64(3) & 16.97(3) & 16.08(6) & 15.76(6) & 15.28(7) \\
\hline
\end{tabular}
\end{center}
\label{Tab:Magnitudes}
\end{table*}

\subsection{X-rays and UV data}
\label{Sec:XrayData}

On 2021 April 06 we performed a 1.5 ks observation of BW~Scl with the Neil Gehrels Swift Observatory \citep{Swift},
using both the X-ray Telescope (XRT; \citealt{SwiftXRT}) and the UV/Optical Telescope (UVOT; \citealt{SwiftUVOT}).
The data were reduced and analysed using {\sc heasoft} 6.28, together with the most recent version of the calibration
files. The UVOT observation was carried out in all six available filters.

{\it Swift}-XRT detected a weak X-ray source with a count-rate of 0.023$\pm$0.004 counts s$^{-1}$.
The spectrum consists of only 35 counts; therefore, no meaningful spectral analysis is possible.
To estimate the X-ray flux of BW~Scl, we assumed that its spectrum is similar to other WZ~Sge-type stars
such as GW~Lib and SSS~J122221.7-311525 \citep{NeustroevSSSX}. By using the count-rate as the scale factor, we
obtain the unabsorbed X-ray flux of BW~Scl in the 0.3--10 keV energy range to be $\sim$7.8\tim{-13} \ergs.
A formal fit of the spectrum with an optically thin emission component (the \apec\ model in \xspec), absorbed by
a variable column gives a consistent estimate of the unabsorbed X-ray flux to be (8.3$\pm$1.6)\tim{-13} \ergs.
This corresponds to the unabsorbed X-ray luminosity of (8.6$\pm$1.7)\tim{29} erg s$^{-1}$, which is in agreement
with the luminosities found for other WZ Sge-type stars in quiescence \citep{Reis13,NeustroevSSSX}.

In order to extend the wavelength coverage of BW~Scl for our analysis, we have extracted the UV data obtained by the \HST\ and
the \emph{Galaxy Evolution Explorer} (\GALEX) satellite. Far-UV echelle spectroscopy
was obtained with the Space Telescope Imaging Spectrograph (STIS) aboard \HST\ in 1999 \citep[for detail, see][]{Gaensicke05}. We
retrieved the spectrum covering the wavelength range of 1150--1710~\AA\ from the StarCAT catalogue \citep{StarCAT}.
\GALEX\ has observed BW~Scl over five separate visits in the NUV (1 in 2003 and 4 in 2005) and four visits in the FUV (1 in 2003
and 3 in 2005). We used the gPhoton package version 1.28.9 \citep{gPhoton} to extract the magnitudes from these observations.

\begin{figure*}
\centering
\resizebox{\hsize}{!}{
\includegraphics[angle=0]{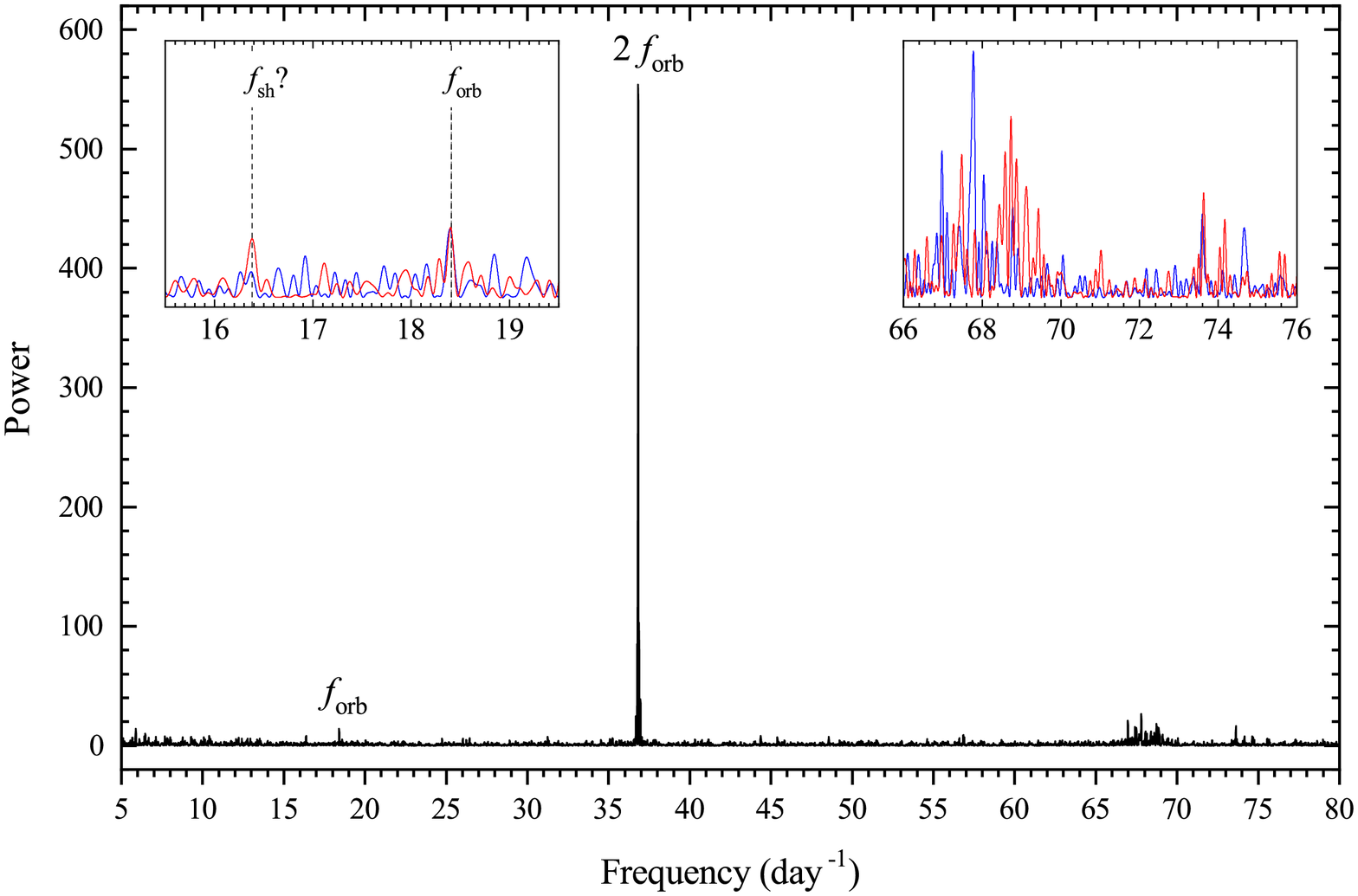}
\includegraphics[angle=0]{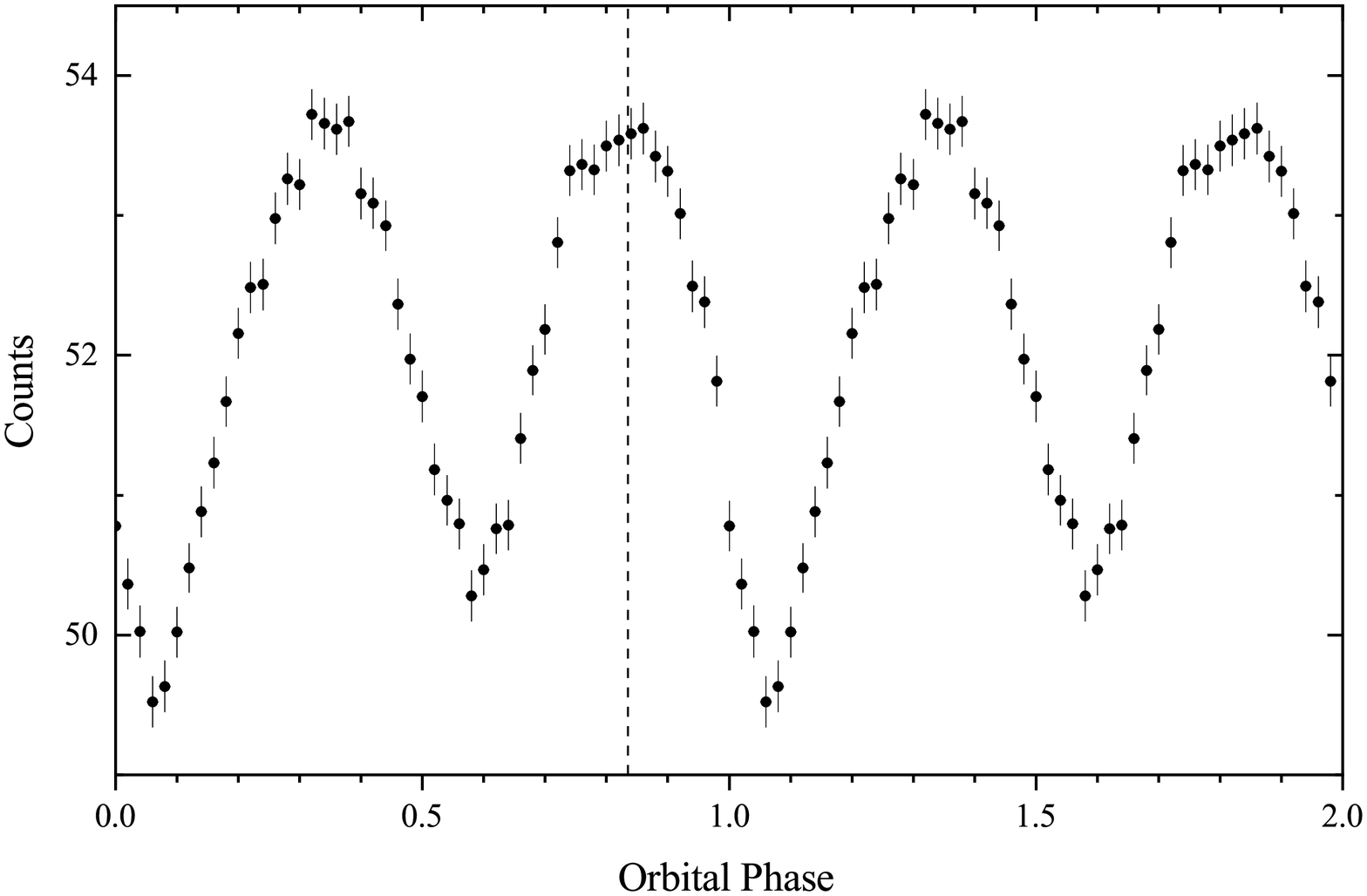}
}
\caption{
Left: Lomb–Scargle power spectrum of the {\it TESS} photometry of BW~Scl. The inset plots show the enlarged regions around
the orbital and 66--76 \cd\ frequencies, in which \citet{Uthas12} have detected transient signals. The blue and red lines in
the insets correspond to the power spectra calculated for the {\it TESS} data obtained before and after the gap, respectively.
Right: the {\it TESS} light curve folded with the $P_{\rm orb}$ according to spectroscopic ephemeris~(\ref{Eq:Ephemeris}) and
averaged in 50 phase bins. The dashed line indicates phase zero according to photometric ephemeris~(\ref{Eq:EphemerisPhot}).
}
\label{Fig:TESS}
\end{figure*}

\begin{figure}
\centering
\resizebox{\hsize}{!}{
\includegraphics[angle=0]{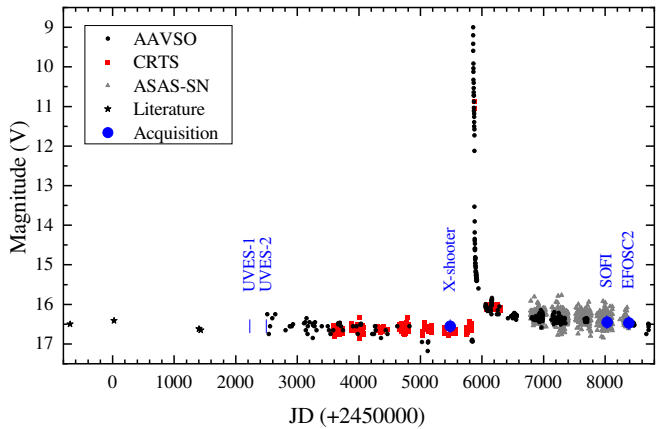}
}
\caption{
Light curve of BW Scl combined of AAVSO, CRTS, and ASAS-SN observations from 1993 up to 2019. Also shown
are measurements found in the literature \citep{Abbott97,Augusteijn97,Uthas12,Gaensicke05}.
The large blue dots represent averaged magnitudes of the object during our observations.
}
\label{Fig:TotalLightCurve}
\end{figure}

\begin{figure*}
\centering
\resizebox{\hsize}{!}{
\includegraphics[angle=0]{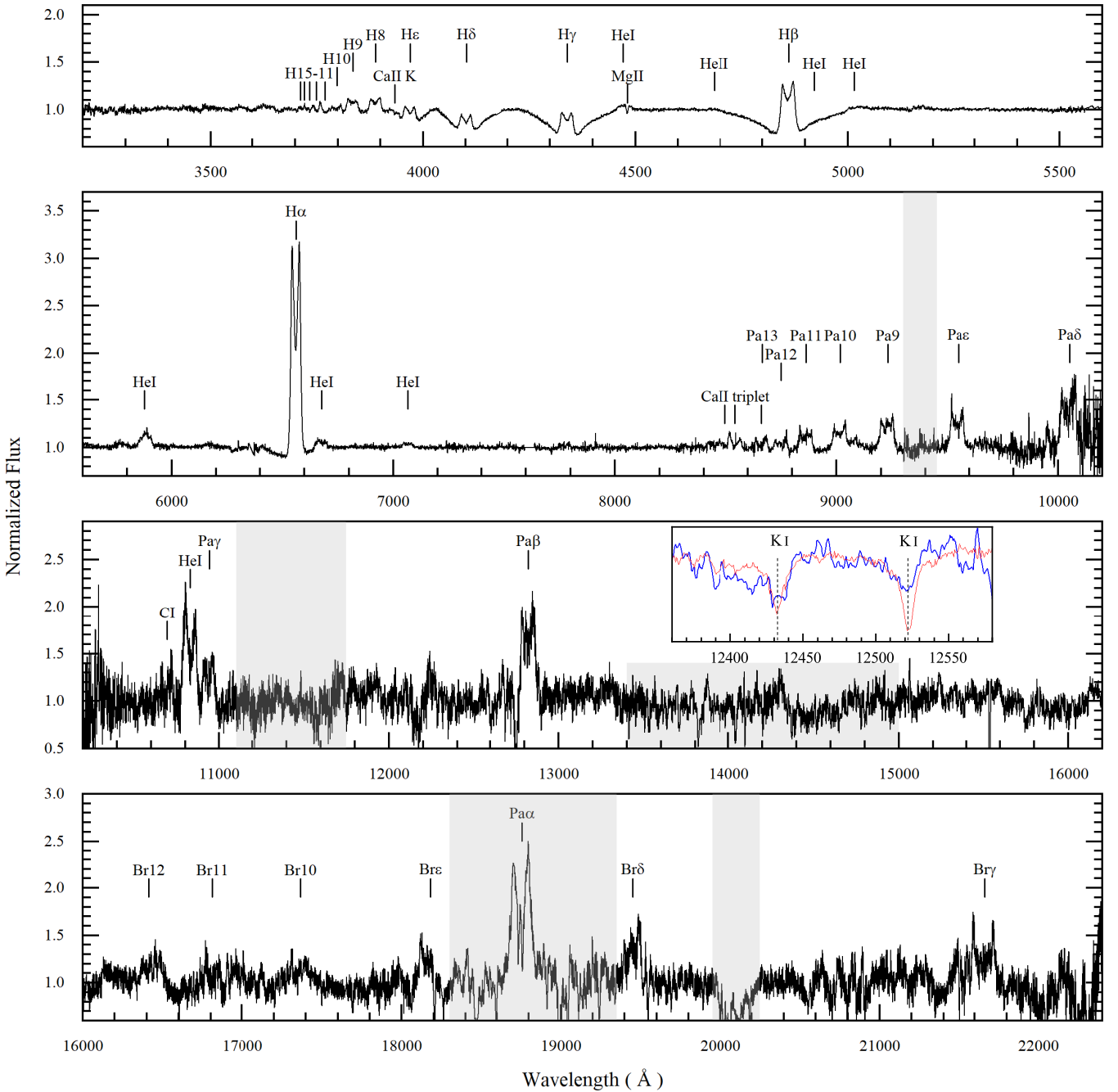}
}
\caption{
Normalized average spectrum of BW~Scl obtained with X-shooter. The most prominent lines are labelled.
The regions contaminated by broad telluric absorption bands are highlighted in grey. The inset shows
the \KI\ absorption lines, which showed up after the individual spectra of BW~Scl were Doppler-corrected
into the frame of the donor star (blue line). The red line is the spectrum of the L9 brown dwarf
DENIS-P J0255.0$-$4700.
}
\label{Fig:Xsh-spec}
\end{figure*}

\begin{figure*}
\centering
\resizebox{\hsize}{!}{
\includegraphics[angle=0]{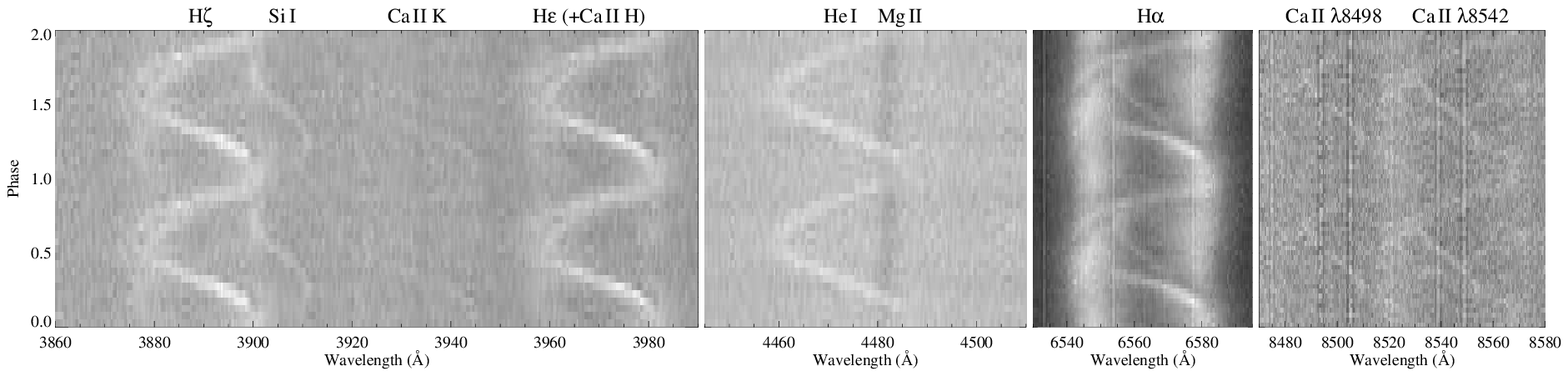}
}
\caption{
Trailed spectra of selected spectral lines from the X-shooter data set, displayed twice for clarity.
The shown lines are dominated by one of, or a mixture of emission and absorption components from the
irradiated donor star (\SiI\ 3906, \CaII\ 8498 and 8542, \Halpha), the hotspot (\Halpha, \Hepsilon, \Hzeta,
\HeI\ 4471, \CaII\ 3934 and 3968), the accretion disc (\Halpha, \Hepsilon, \Hzeta, \CaII\ 8498, 8542), and
the WD (\MgII\ 4481, \CaII\ 3934). White indicates emission shown in a linear scale.
}
\label{Fig:TrailedLines}
\end{figure*}

\section{Photometric variability of BW~Scl}

\subsection{Short-term variability}
\label{Sec:TESS}

BW~Scl in quiescence is known to exhibit short-term photometric variability with typical amplitude of $\sim$0.1 mag
\citep{Uthas12}. Its light curve is dominated by cyclic modulations at half the orbital period \citep{Augusteijn97},
for which \citet{Uthas12} presented an ephemeris based on photometric observations from the years 1995--2009. The ephemeris is
thus covering not only the UVES and X-shooter observations, but is also expected to be valid for our NTT data. Indeed, Doppler
maps which utilize these data and calculated according to the period from \citet{Uthas12} exhibit no obvious phase difference
(see Section~\ref{Sec:DopMap}). However, phase zero of the ephemeris causes confusion. According to \citet{Uthas12}, it
corresponds to ``orbital maximum'', whereas phase zero in their figure 6 corresponds to one of the (not deepest) light minima.

BW~Scl was observed with the {\it Transiting Exoplanet Survey Satellite} ({\it TESS}) from 2020 August 26 to September 20 in Sector
29 at 20 second cadence. These observations provide the opportunity to obtain a long continuous coverage to test the stability of
double-humped modulations in BW~Scl, to resolve the confusion regarding phase zero, and to study other variabilities. We downloaded
the {\it TESS} data in the Pre-search Data Conditioning Simple Aperture Photometry flux (PDCSAP) format from the Mikulski Archive
for Space Telescopes (MAST) archive. In the PDCSAP light curves, instrumental systematic variations have been removed from the data
in the pipeline reduction procedure. The light curve has a gap between 2020 September 6.23--9.83 (UT).

We calculated Lomb--Scargle periodograms for the total light curve (\autoref{Fig:TESS}, left-hand panel), and separately for its
two pieces consisting of the data obtained before and after the gap. The periodograms are exclusively dominated by a very strong
and sharp peak at half the orbital frequency of 36.816$\pm$0.015 d$^{-1}$, consistent with the findings of \citet{Uthas12},
although a signal at the orbital frequency is also clearly present. \autoref{Fig:TESS} (right-hand panel) shows the {\it TESS}
light curves folded with the orbital period and averaged in 50 phase bins. The folded light curve exhibits double-humped
modulations with a total amplitude of $\sim$0.08 mag and a small even–odd asymmetry.
Attempting to improve the ephemeris, we combined the {\it TESS} data with the observations from the AAVSO (American Association of
Variable Star Observers) obtained in 2002--2020, and published times of maxima from \citet{Augusteijn97}. A linear least-square
fit to the O--C values yielded the linear heliocentric ephemeris of the light maxima:
    \begin{equation}  \label{Eq:EphemerisPhot}
        T_{\mathrm{max}}=\textnormal{HJD } 245\,0032.18165(26) + 0.0543239136(24) \times E \,.
    \end{equation}
The new value of the orbital period is very close to that of \citeauthor{Uthas12}'s ephemeris and results in just a 0.02 phase
shift over 25 yr. Below, in Section~\ref{Sec:SysPar}, we find the time of the inferior conjunction of the donor star using
the X-shooter spectra. This enables us to define an accurate spectroscopic ephemeris, in which phase zero corresponds to the
inferior conjunction of the donor:
    \begin{equation}  \label{Eq:Ephemeris}
        T_{0}=\textnormal{HJD } 245\,0032.13631(11) + 0.0543239136(24) \times E \,.
    \end{equation}
Thus, the primary light maximum from the ephemeris~(\ref{Eq:EphemerisPhot}) occurs at spectroscopic phase 0.864 (\autoref{Fig:TESS},
right).

In addition to the double-humped modulations, \citet{Uthas12} also reported the detection of a few other photometric variabilities
like complex signals with a period near 20~min ($\sim$68 \cd) and a transient `quiescent superhump' signal with a period of
$P_{\rm sh}$= 87.27~min. Based on the {\it TESS} data, we confirm the presence, in the power spectrum, of multiple peaks with
periods around 20~min. However, the data obtained before and after the gap show different sets of peaks in this frequency range
(see the right inset plot in the left-hand panel of \autoref{Fig:TESS}). A `superhump' signal is also present (it is much stronger
in the `after the gap' data where it reaches a high enough confidence level, see the left inset plot in the left-hand panel of
\autoref{Fig:TESS}) but with a bit different frequency of 16.38 \cd\ ($~\sim$87.9 min).

\subsection{Long-term photometric evolution}
\label{Sec:LongTermPhot}

On the long-term time-scale of years the average brightness of BW~Scl appears to be quite stable (\autoref{Fig:TotalLightCurve}).
However, the light curve composed of the photometric observations from the AAVSO, CRTS \citep[Catalina Real-Time Transient Survey;][]{CRTS},
and ASAS-SN \citep[The All-Sky Automated Survey for Supernovae;][]{ASASSN1,ASASSN2} shows that ten years after
the superoutburst in 2011, BW~Scl has not yet returned to the preoutburst level. Our own photometric measurements
indicate that during the EFOSC2 and SOFI observations the object has been $\sim$10 percent brighter than during
the X-shooter observations.

We also compared the available UV and IR observations obtained between 1999 and 2021 with different telescopes to check for
possible variability. We found an excellent agreement between the IR fluxes measured by the \emph{Spitzer Space Telescope}
in 2004 \citep{Spitzer} and by the \emph{Wide-field Infrared Survey Explorer} in 2010 \citep[the \WISE\ mission;][]{WISE} and in 2017-2020
\citep[the \NEOWISE\ mission;][]{NEOWISE,NEOWISER}.
All the GALEX measurements obtained in 2003 and 2005 also show no sign of variability with the standard deviation
of the data in each band $<$0.05 mag, and they are highly consistent with the UVOT observations obtained in 2021 and the \HST\
spectroscopy obtained in 1999. Thus, we conclude that during at least 12 yr before the superoutburst (and probably a longer
time), BW~Scl was at the same flux level.

\begin{figure}
\resizebox{\hsize}{!}{
\includegraphics{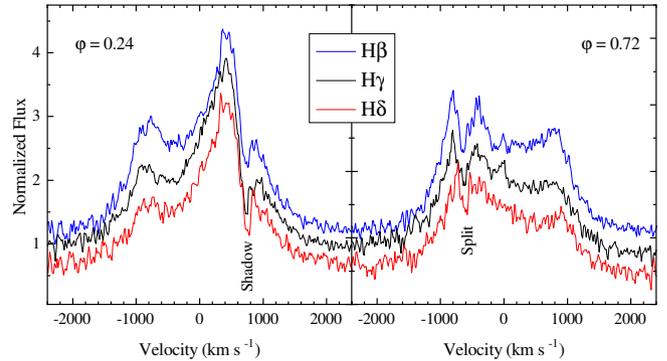}
}
\caption{The emission line profiles of \Hbeta, \Hgamma, and \Hdelta, obtained with X-shooter at phases 0.24 (left) and 0.72
(right). The profiles show the S-wave shadow and split, correspondingly. The profiles shifted vertically for clarity.
}
\label{Fig:ProfileFeatures}
\end{figure}

\section{Averaged and trailed spectra of BW Scl}
\label{Sec:AverageSpec}

\autoref{Fig:Xsh-spec} shows the normalized average X-shooter spectrum of BW~Scl, covering a larger wavelength range
and having the highest SNR in the optical domain of all the available data. We note, however, that although all the
spectra are similar in appearance, they exhibit a notable difference in the strength of emission lines. In the
X-shooter spectrum they are $\sim$2 times less intense than in other data sets (\autoref{Fig:SpecOpt-Comp}).

The blue part of the BW~Scl spectrum is dominated by broad Balmer absorption lines from the WD with superimposed double-peaked
emission components originated in the accretion disc. Balmer emissions can be traced up to H14. Weak emission lines of \HeI\
and \CaII\ 3968~\AA\ are also present. The \MgII\ 4481~\AA\ absorption line from the WD is clearly visible. At longer
wavelengths, besides the \Halpha\ and \HeI\ emission lines, double-peaked hydrogen emission lines of the Paschen and Brackett
series are present (can be traced from \Paalpha\ up to Pa14 and from \Brgamma\ up to Br13, respectively). Additionally, we
detect the \CaII\ triplet (8498, 8542, and 8662~\AA) and, possibly, a blend of \CI\ lines at $\sim$10693~\AA, both in emission.

We note that in contrast to the hydrogen \emph{double-peaked} emission lines, all the \HeI\ lines except for 10830 \AA\
exhibit, in the averaged spectra, broad smoothed profiles. The reason for it, as shown below, is that the accretion disc
appears to be not hot enough to excite helium atoms so as to produce prominent \HeI\ lines. Instead, the \HeI\ emission
comes primarily from the hotspot, the area of interaction between the gas stream and the disc, where the temperature is
sufficient.

This and other important details can be found in the phase-resolved trailed spectra (\autoref{Fig:TrailedLines}).
It is seen that many lines consist of a mixture of several emission and absorption components originated in the disc,
the hotspot, the irradiated donor star, and/or the WD. For instance, all the hydrogen lines (\Halpha, \Hepsilon, and
\Hzeta\ are shown in the figure) in addition to the double-peaked emission from the disc also exhibit an S-wave moving
in between. This S-wave which originates in the hotspot is the only visible emission component in all the \HeI\ lines
except for 10830 \AA. Surprisingly, the hotspot S-wave is even visible in the high excitation line of \HeII\ 4686 \AA,
although it is not detected in the Bowen blend. Similar to \HeII\ 4686 \AA, only the hotspot produces a weak
\OI\ 7774~\AA\ emission and there is no sign neither of the  disc nor the donor in this line. The \NaI\ doublet
around 8190 \AA\ which is a very powerful diagnostic for the donor star in CVs \citep{Friend88} is not detected at all.

The Balmer trailed spectra (\autoref{Fig:TrailedLines}) and even individual Balmer profiles (\autoref{Fig:ProfileFeatures})
exhibit interesting features of the S-wave. The S-wave apparently splits between phases 0.5 and 0.9 when it is blue-shifted,
and between phases 0.1 and 0.5 a sharp absorption shadow appears which precedes the S-wave. A similar `S-wave shadow' has
been reported by \citet{SpruitRutten} for the prototype object WZ~Sge.\footnote{Although \citet{SpruitRutten} did not detect
the S-wave splitting, the latter is detectable in the \Halpha\ and \Hbeta\ trailed spectra of WZ~Sge presented by \citet{WZ1},
see their fig.~1.} We discuss these phenomena in Section~\ref{Sec:DopMap}.

Of special interest is the detection in \Halpha\ of an additional S-wave with a smaller radial velocity amplitude, which
is phase-shifted with respect to the hotspot S-wave. Similar features also detected in
a few other lines such as \FeI\ (5270, 5329, 5372~\AA) and the \CaII\ triplet (8498, 8542, and 8662~\AA), but it is most
clearly seen in \SiI\ 3906 \AA\ (\autoref{Fig:TrailedLines}). This sharp emission feature becomes stronger between phases
0.2--0.8 and disappears between phases 0.8--0.2, moving in antiphase with the narrow absorption lines from the WD (in
addition to \MgII\ 4481 \AA, such an absorption is also visible in \CaII\ 3934 \AA). All these properties are consistent
with being produced by the inner face of the donor companion to the WD. We note that the \SiI\ 3906 \AA\ line is rarely
seen in spectra of CVs, but has been detected in some pre-CV objects \citep[see e.g.][]{Parsons11}.

\begin{table}
\caption{Elements of the radial velocity curves of the donor star and the WD derived from the X-shooter spectra.
}
\begin{tabular}{lclc}
\hline\hline
\noalign{\smallskip}
Spectral lines        &    $K$          &  $\gamma$-velocity    & $\phi_{0}$ \\
                      & (\kms)          &       (\kms)          &  \\
\noalign{\smallskip}
\hline
\noalign{\smallskip}
\multicolumn{3}{l}{Irradiation driven emission lines from the donor:} \\
\SiI\ 3906 & 410.0$\pm$3.6   &     -4.8$\pm$3.5      & 0.500$\pm$0.002 \\
\FeI\ 5270+5328+5371  & 415.9$\pm$3.9   &    -3.4$\pm$3.0      & 0.505$\pm$0.002 \\
\CaII\ triplet        & 402.6$\pm$2.3   &    -1.9$\pm$2.2      & 0.501$\pm$0.002 \\
\Halpha\ $^a$         & 422.9$\pm$7.4   &    -2.9 (fixed)      & 0.504$\pm$0.004 \\
All lines             & 405.5$\pm$1.4   &    -2.9$\pm$1.4      & 0.502$\pm$0.001 \\
\noalign{\smallskip}
\hline
\noalign{\smallskip}
\MgII\ 4481 from the WD: & 27.7$\pm$3.0 &    53.0$\pm$2.1      & -0.003$\pm$0.012 \\
\noalign{\smallskip}
\hline
\noalign{\smallskip}
\end{tabular}\\
$^a$ Fitting of the \Halpha\ radial velocities was performed using the fixed $\gamma$-velocity.
\label{Tab:RadVel}
\end{table}

\section{Determination of system parameters}
\label{Sec:DetSysPar}

\subsection{Radial velocities of the donor star}

\begin{figure}
\centering
\resizebox{\hsize}{!}{
\includegraphics[angle=0]{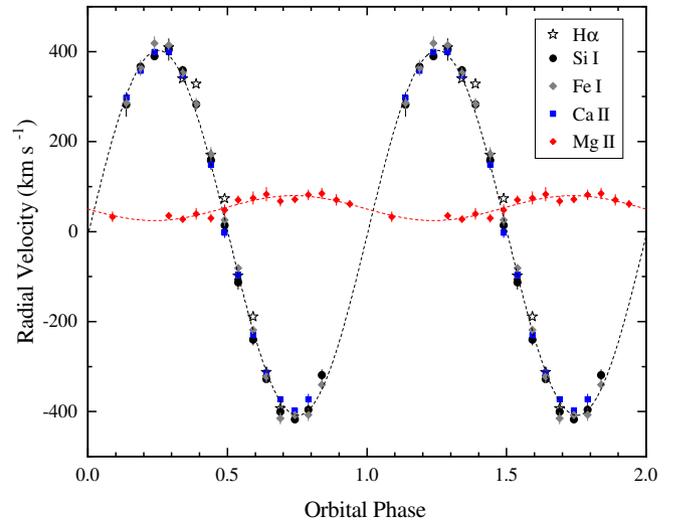}
}
\caption{
Radial velocity fits to the \MgII\ absorption line from the WD (red diamonds) and emission lines from the donor star (other symbols)
in BW~Scl.
}
\label{Fig:RadVel}
\end{figure}

The X-shooter spectra show several irradiation driven emission lines which can be used to measure the radial velocity semi-amplitude
of the donor star. \SiI\ 3906 \AA\ is the most clearly visible such a line. It is well detectable in phase-folded spectra manifesting
itself as a narrow and sharp peak (\autoref{Fig:TrailedLines}). The \CaII\ triplet and \FeI\ lines are noisier than
\SiI\ which makes it difficult to obtain reliable measurements of radial velocities. To suppress noise, we converted the wavelengths
to velocities for each of the line components and combined their fluxes together, separately for \CaII\ and \FeI. \Halpha\ is another
line in which the irradiation-driven emission component is clearly visible, although it is heavily contaminated by other
sources of emission, especially at the phases prior to 0.28. We employed a Gaussian fit to measure the radial velocities of the peaks
in \SiI\ and \Halpha\ and the combined \CaII\ and \FeI\ lines. The resulting radial velocity curves were fitted, separately for \SiI,
\CaII, and \FeI\ with a sinusoid of the form
    \begin{equation}  \label{Eq:radvelfit}
      V(\varphi)=\gamma -K \sin \left[ 2\pi \left(\varphi-\varphi_0 \right) \right] \,.
    \end{equation}
We obtained very consistent values for the systemic velocity $\gamma$ and the phase zero-point $\phi_0$, and slightly different for
the radial velocity semi-amplitude $K_2$=$K$ (\autoref{Tab:RadVel}). A limited amount of measurements of \Halpha\ does not
allow constraining the $\gamma$-velocity, but by fixing $\gamma$ at the average value of $-$2.9~\kms\ we found that the measurements
of the \Halpha\ line and the calculated parameters are very consistent with other lines (see \autoref{Fig:RadVel} and
\autoref{Tab:RadVel}).

Although the difference in the $K_2$ values can be real
(these lines originate on the inner hemisphere of the donor star, thus their radial velocities do not track the star's
centre-of-mass), we combined all the measurements together and fitted them again to obtain the final values of
$K_2$=405.5$\pm$1.4 \kms\ and $\gamma=-2.9\pm$1.4 \kms.

\begin{figure}
\centering
\resizebox{\hsize}{!}{
\includegraphics[angle=0]{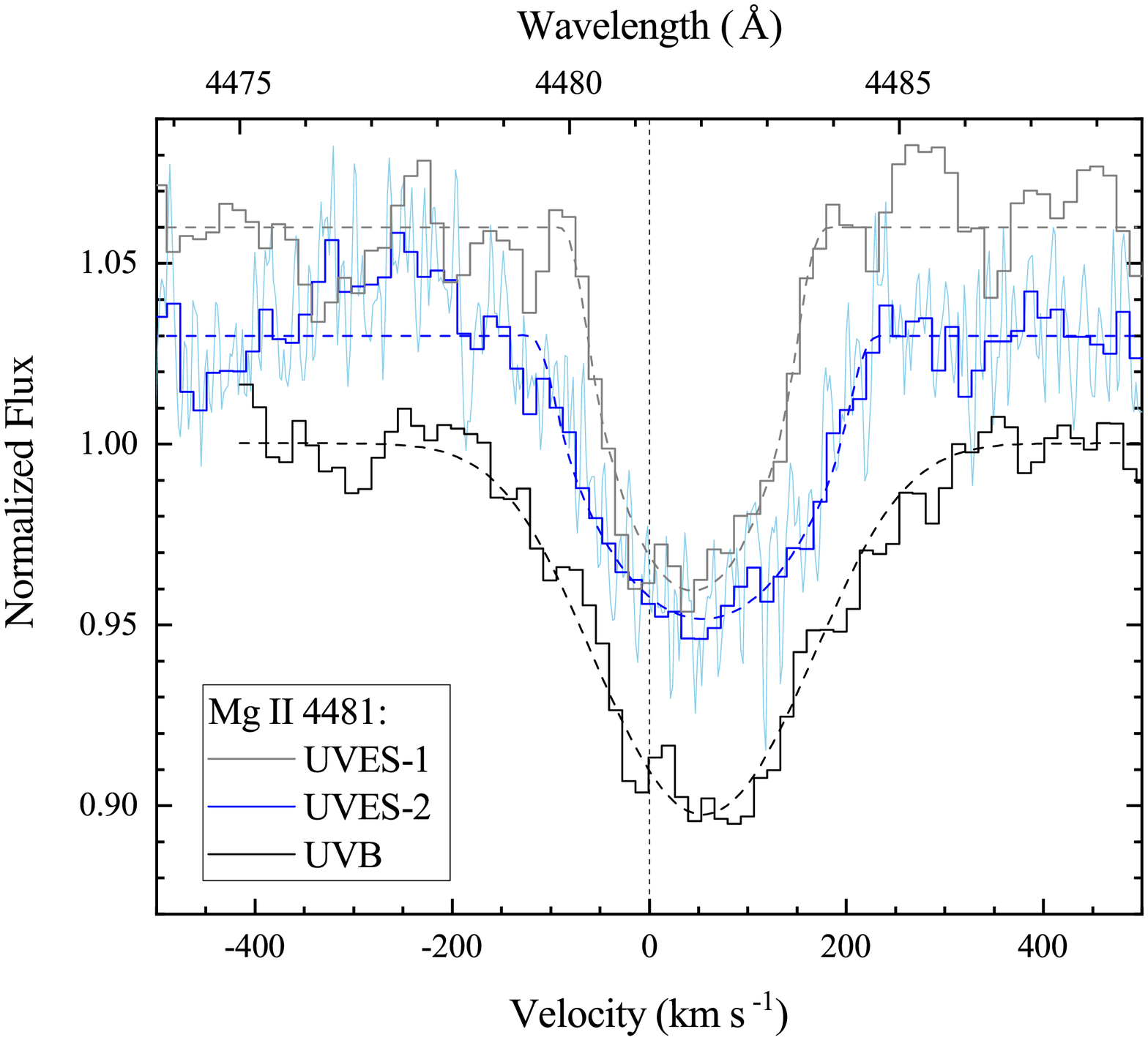}
}
\resizebox{\hsize}{!}{
\includegraphics[angle=0]{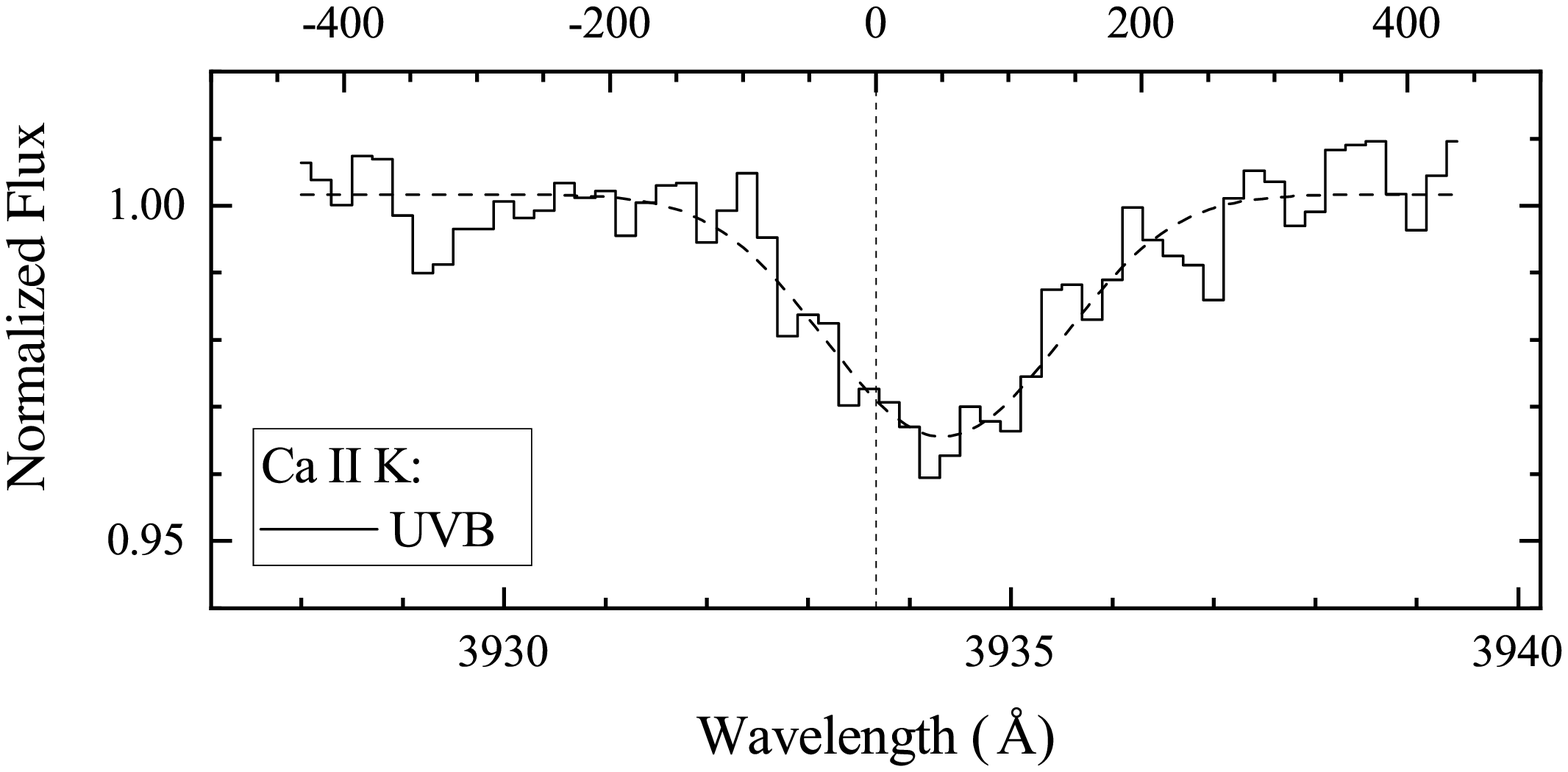}
}
\caption{The averaged \MgII\ line from different data sets normalized to the continuum and shifted upward by 0.03 units
(upper panel), and the \CaII~K line from the X-shooter spectrum (bottom panel).
For display purposes, the noisy UVES spectra were rebinned to the 0.2~\AA\ wavelength step, although the original UVES-2
spectrum is also shown as the light-blue line. The best-fitting rotational broadening profiles are overplotted as dashed lines.
The dotted vertical lines indicate the laboratory wavelengths of 4481.21 and 3933.66~\AA, respectively.}
\label{Fig:MgII}
\end{figure}

\subsection{Rotational and radial velocities of the WD}
\label{Sec:MgII}

Narrow absorption lines originated in the photosphere of the WD in BW~Scl provide an important tool for measuring radial velocity
modulations of the WD and of its rotational velocity. Moreover, these lines are subject to gravitational redshift allowing direct
measurement of the WD mass \citep[see e.g.][and references therein]{vanSpaandonk10}. While the \CaII~K absorption line is quite
weak and can be detected in the X-shooter spectra only (\autoref{Fig:TrailedLines}), the \MgII\ 4481~\AA\ line is strong enough
to be visible in both the average UVES and X-shooter data sets. Still,  the weakness of the \MgII\ line coupled together with the low
SNR of the UVES spectra prevented us from obtaining realistic measurements of the radial velocities using the UVES data. To measure
$K_{1}$, we used the phase-folded X-shooter spectra only, in which radial velocity modulations of \MgII\ are recognizable even
with the naked eye. We note that between the orbital phases $\sim$0.9 and $\sim$1.2 the \MgII\ absorption is contaminated by
the \HeI\ 4471\,\AA\ hotspot emission. For this reason, the corresponded spectra were excluded from the analysis.
We followed an interactive approach from \citet{BF_Eri}:

\begin{enumerate}
 \item At the first step, we obtained the cross-correlation template spectrum, which was used for measuring the radial velocities
 of the \MgII\ 4481\,\AA\ line. Initially, the template was obtained by simple averaging of all the UVB spectra.

 \item  \label{MgII:step2}
 Next, the phase-folded spectra were cross-correlated with the template, using the wavelength region 4470--4495 \AA.
 The resulting radial velocity curve was then fitted with the expression~(\ref{Eq:radvelfit}), and preliminary values of
 $K_1$=$K$, the $\gamma$ velocity, and the phase zero-point $\phi_0$ were calculated.

 \item  \label{MgII:step3}
 Each spectrum in the X-shooter data set was then shifted to correct for the orbital motion of the WD, and the results averaged,
 creating thus a new template.

 \item Step~\ref{MgII:step2} was then repeated and the final values of the parameters were calculated which are summarized in
 \autoref{Tab:RadVel}. \autoref{Fig:RadVel} shows the measured radial velocities together with their sinusoidal fit. Note that
 the difference between the phase zero-points obtained from the emission and absorption lines is very close to 0.5, as it must
 be if the derived velocities from those lines trace the components' motion.

 \item Finally, we repeated step \ref{MgII:step3} again, now for all the X-shooter and UVES data sets, using the adopted
 values of $K_1$ and $\phi_{0}$ but leaving $\gamma$=0. The obtained average profiles of \MgII\ (and \CaII~K
 in the X-shooter spectrum) were used to derive the mean radial velocity of the WD and its rotational velocity.
\end{enumerate}

We find that the appearance of the \MgII\ profile is slightly different in the obtained average spectra (\autoref{Fig:MgII},
upper panel). The line is less strong
in the UVES-2 spectrum, although its FWHM is similar to that of the UVB spectrum. The line in UVES-1 appears more narrow than in
the other two spectra. However, this spectrum has a much lower SNR (\autoref{Tab:MgII}) which may affect the line appearance.
The \MgII\ equivalent width (EW) is the largest in the UVB spectrum. This suggests that the \MgII\ abundance is variable in the
WD atmosphere \citep{Holberg97}.

To derive the mean radial velocity of the WD, $\upsilon_{\rm rad}$, and its rotational velocity, \vsini, we fitted the
Doppler-corrected co-added \MgII\ and \CaII~K profiles with a synthetic profile which was smoothed to the corresponded spectral
resolution of the data set and rotationally broadened using the \PyA\footnote{\url{https://github.com/sczesla/PyAstronomy}} function
{\sc ROTBROAD} \citep{pyAst}.
Because the \MgII\ 4481 \AA\ line is a triplet with the components at 4481.126, 4481.150, and 4481.325~\AA\ whose transition
probabilities are $\log$($gf$) = 0.7367, $-$0.5643, and 0.5818, respectively, for the fitting we used a spectral model based
on a list of Gaussian lines from {\sc PyAstronomy's model suite (PYASTRONOMY.MODELSUITE.LLGAUSS)}.
The rotational broadening solution depends upon the limb-darkening which, to the best of our knowledge, has never been evaluated
for metal spectral lines from WDs. However, the limb-darkening coefficients have been computed for a number of broad-band photometric
filters \citep[see e.g.][]{LimbDark}. For the range of expected parameters of the DA WD in BW~Scl ($T_{\rm eff}\sim$12000--15000~K,
$\log g\sim$8.25--8.75), the linear limb-darkening coefficient $\varepsilon$ ({\sc PYASTRONOMY} uses a linear approximation) in the
$B$--band varies between $\sim$0.32 and 0.38. We assumed $\varepsilon$ = 0.35 in deriving the rotational broadening. All fits yield
fairly consistent results but with slightly different values of the rotational velocity from the \CaII\ and UVES-1 \MgII\ profiles
(\autoref{Tab:MgII}). As it has been shown by \citet{vanSpaandonk10}, the low SNR of the spectrum and the line weakness can affect
the determination of the model parameters. Using the SNR as a weighting factor, we calculated the averaged values of
$\upsilon_{\rm rad}$=54.6$\pm$3.0 \kms\ and \vsini=139$\pm$6 \kms. The found rotational velocity of the WD in BW~Scl is
consistent with \citet{Gaensicke05}, but it appears relatively slow in comparison with other accreting WDs \citep{SionGodon22}.

\begin{table}
\begin{center}
\caption{Measurements of the \MgII\ 4481 \AA\ and \CaII~K lines.}
\begin{tabular}{lcccccc} \hline \hline
Data set   &  SNR   &  EW   &  FWHM  & Line     &$\upsilon_{\rm rad}$ &   \vsini     \\
           &        & (\AA) &  (\AA) & strength &     (\kms)           &   (\kms)    \\
\hline
UVES-1     &   30   & 0.30  &  2.9   & 0.90     &    44.3$\pm$4.2     &  117$\pm$5   \\
UVES-2     &  105   & 0.27  &  3.6   & 0.92     &    53.8$\pm$3.4     &  151$\pm$5   \\
UVB (\MgII)&  165   & 0.38  &  3.6   & 0.90     &    55.2$\pm$2.3     &  140$\pm$6   \\
UVB (\CaII)&  140   & 0.07  &  2.0   & 0.96     &    55.6$\pm$4.5     &  101$\pm$15  \\
\hline
average    &        &       &        &          &    54.6$\pm$3.0     &  139$\pm$6   \\
\hline
\end{tabular}
\label{Tab:MgII}
\end{center}
\end{table}

\subsection{The mass of the WD from the gravitational redshift}
\label{Sec:GravRedShift}

The observed \MgII\ and \CaII\ velocity $\upsilon_{\rm obs}$=$\upsilon_{\rm rad}$=54.6$\pm$3.0 \kms\ is a sum
of the systemic velocity $\gamma$ and the gravitational redshift $\upsilon_{\rm grav}$:
\begin{equation}
\upsilon_{\rm obs}=\gamma + \upsilon_{\rm grav} \,.
\end{equation}
$\gamma$ is most accurately measured from radial velocity variations of the donor star (the contribution of the low-mass secondary
to the redshift is negligible). Adopting the \SiI+\CaII+\FeI\ value $\gamma$=$-$2.9$\pm$1.4 \kms\ from \autoref{Tab:RadVel},
we obtain $\upsilon_{\rm grav}$ = 57.5$\pm$3.3 \kms. The gravitational redshift depends on the ratio between mass and radius
\citep{GreensteinTrimble}:
\begin{equation}
{\upsilon_{\rm grav}} =\frac{G M}{c R} = 0.637\, \frac{M_{\rm WD}}{M_\odot} \frac{R_\odot}{R_{\rm WD}}\,\mathrm{km\,s}^{-1} \,,
\end{equation}
where $G$ is the gravitational constant, $c$ is the speed of light, and $R_{\rm WD}$ and $M_{\rm WD}$ are the radius and the mass
of the WD, respectively. Using Nauenberg's analytic mass--radius relation for WDs \citep{Nauenberg,CookWarner84}, we
get $M_{\rm WD}$=0.87$\pm$0.03~M$_{\sun}$. However, this value needs to be lowered since Nauenberg's relation assumes a cold,
non-rotating WD, providing thus only a lower limit to the WD radius for a given mass. The radius of a WD at a given temperature
and mass depends on the thickness of the hydrogen layer at the WD surface \citep{Romero19}. According to \citet{Bedard20}, a WD
with the mass of 0.8--1.0 \Msun\ and the temperature of $\sim$15000~K has the radius about 2 per cent larger than that of a cold
WD in case of a thin hydrogen layer, and about 3 per cent in case of a thick layer.
Because the thickness of the hydrogen layer is not known, in the following we adopt the average conservative value
$M_{\rm WD}$=0.85$\pm$0.04~M$_{\sun}$, which corresponds to the surface gravity of $\log g$=8.40$\pm$0.02.
The found mass of the WD in BW~Scl is consistent with the mean WD mass in CVs of $0.81^{+0.16}_{-0.20}$~\Msun\
\citep{Pala22,Zorotovic11} but appears less than that deduced by \citet{Pala22} from the UV spectral fit (1.007$\pm$0.012~\Msun).

\subsection{System parameters}
\label{Sec:SysPar}

The detection of radial velocity variations of both stellar components and a direct measurement of the WD mass
via gravitational redshift enable us to fully determine the orbital and stellar parameters of the binary. The
ratio of the radial velocity semi-amplitudes of the primary and secondary stars $K_1$ and $K_2$ gives the mass
ratio $q$
   \begin{equation}
    \label{Eq:q}
      q = {M_2 \over M_1} = {K_1 \over K_2}\,.
   \end{equation}

\noindent
Combining the values of $K_1$, $K_2$, $P_{\rm orb}$, and $M_1$=\Mwd\ we can find the system inclination
$i$, the mass of the donor $M_2$, and the binary separation $a$:
   \begin{equation}
    \label{Eq:M1sini}
      M_1 \sin^3 i = {P K_2 (K_1 + K_2)^2 \over 2 \pi G} \,,
   \end{equation}
   \begin{equation}
    \label{Eq:M2sini}
      M_2 \sin^3 i = {P K_1 (K_1 + K_2)^2 \over 2 \pi G} \,,
   \end{equation}
   \begin{equation}
      a \sin i = {P (K_1 + K_2) \over 2 \pi} \,.
   \end{equation}

\noindent
Here we have to note that the \emph{observed} value of $K_2$=$K_{\rm 2,o}$ represents only a lower limit to the
true radial velocity semi-amplitude $K_{\rm 2,true}$ because the detected emission lines are irradiation-induced
from the surface of the donor star facing the WD. The non-coincidence of the
centre-of-mass and the centre-of-light of the donor results in systematic errors in the determined $K_{\rm 2,o}$.
Therefore, in order to obtain the true value of $K_2$, a $K$-'correction' should be applied \citep{WadeHorne88}.
The $K$-correction can be expressed as:
\begin{equation}
\label{Eqn:Kcor}
 \frac{K_{2,\rm o}}{K_2} \approx 1-0.462 q^{1/3} (1+q)^{2/3} f \,,
\end{equation}
\citep[equation 2.77 in][]{Warner}, where $f=\Delta R_{2}/R_{2}$ is the ratio of the displacement of the centre-of-light
from the centre of mass of the secondary star to the radius of the secondary. For $f$=0 the emission is spread uniformly
across the entire surface of the donor, $f$=0.5 roughly corresponds to the extreme case where the spectral line comes only
from the hemisphere closest to the WD, and for $f$=1 all the emission comes from the donor surface near the inner Lagrangian
point. It is difficult to quantify the $f$-factor, but it seems to depend on the optical depths of spectral lines that can
explain the scatter of the measured values of $K_{\rm 2,o}$ for different lines (see also our \autoref{Tab:RadVel}).
According to \citet{Parsons10,Parsons12}, $f$$\approx$0.77 for an optically thin line, and
$f$=0.5 for optically thick emission. Using these values, we obtain $K_{\rm 2,o}$/$K_{\rm 2,true}$=0.88$\pm$0.02 and,
accordingly, $K_{\rm 2,true}$=461$\pm$12, where 0.02 and 12 are not the standard deviations but rather adopted
ranges of values.

Now, from the above equations, we find $q$=0.060$\pm$0.007, $M_2$=0.051$\pm$0.006\,\Msun, $i$=64.3$\pm$3.6\degr, and
$a$=0.584$\pm$0.020\,\Rsun. These and other deduced system parameters are listed in Table~\ref{Tab:Syspar}. The table
also shows the tidal truncation radius of the disc $r_{\rm d,max}$, calculated using updated approximation formula (3) from
\citet{NeustroevHT2}. $r_{\rm d,max}$=0.338\,$R_{\sun}$=0.58\,$a$ corresponds to the minimal possible Keplerian velocity in the disc
of $\upsilon_{\rm out}$ = 693 \kms, and the minimal observed velocity $\upsilon_{\rm min}$=$\upsilon_{\rm out} \sin i$ = 624 \kms.

\begin{table}
\centering
\caption[] {Orbital and system parameters for BW Scl.}
\begin{tabular}{lc}
\hline
\hline\noalign{\smallskip}
Parameter                           & Value based on the          \\
                                    & corrected $K_{\rm 2}$=$K_{\rm 2,true}$ \\
\noalign{\smallskip}
\hline\noalign{\smallskip}

\Porb\ (d)                          & 0.0543239136(24)            \\
$T_{0}$ (HJD)                       & 245\,0032.13631(11)         \\
$K_1$ (\kms)                        &  27.7$\pm$3.0               \\
$K_{\rm 2,o}$ (\kms)                &  405.5$\pm$1.4              \\
$K_{\rm 2,true}$ (\kms)             &  461$\pm$13                 \\
$\gamma$ (\kms)                     &  -2.9$\pm$1.4               \\
$q=M_{2}/M_{1}$                     &  0.060$\pm$0.006            \\
$M_{\rm wd}/M_{\sun}$               &  0.85$\pm$0.04              \\
$M_{2}/M_{\sun}$                    &  0.051$\pm$0.006            \\
$i$                                 &  64.3$^\circ\pm$3.6$^\circ$ \\
$a/R_{\sun}$                        &  0.58$\pm$0.02              \\
$R_{2}/R_{\sun}$ (volume radius)    &  0.103$\pm$0.005            \\
$r_{\rm d,max}/ R_{\sun}$             &  0.338$\pm$0.001            \\

\noalign{\smallskip}
\hline

\end{tabular}
\label{Tab:Syspar}
\end{table}

\section{The WD, donor star, and accretion disc}
\subsection{Temperature of the WD in BW~Scl}
\label{Sec:WDmodel}

The blue spectrum of BW~Scl is dominated by the characteristic broad Balmer absorption lines from the DA WD, which can be used
to evaluate the WD effective temperature $T_{\rm eff}$ and surface gravity $\log g$. These parameters of isolated WDs are often
estimated through a spectral fit of a grid of WD synthetic atmosphere models to the observed absorption lines. In accreting WDs,
the WD spectrum is contaminated by the accretion disc and the donor star, which makes it a more complex task to determine
$T_{\rm eff}$ and $\log g$. While the disc lines can be cut out of the spectrum,
the shape of the disc continuum is still unknown. The latter is often assumed to follow a simple power law, although
it has been shown that in the low mass transfer rate regime the disc emission appears to be better reproduced by a hydrogen slab
model \citep[see, e.g.,][]{HernandezSantisteban16,Pala-QZLib,Pala-SDSS1238,Pala22}. Nevertheless, given that the hydrogen slab
is a multiparameter model and that we include in our analysis only a relatively short portion of the spectrum, in the following
we use a power law as a first approximation to the disc continuum.

\begin{figure*}
\resizebox{\hsize}{!}{\includegraphics{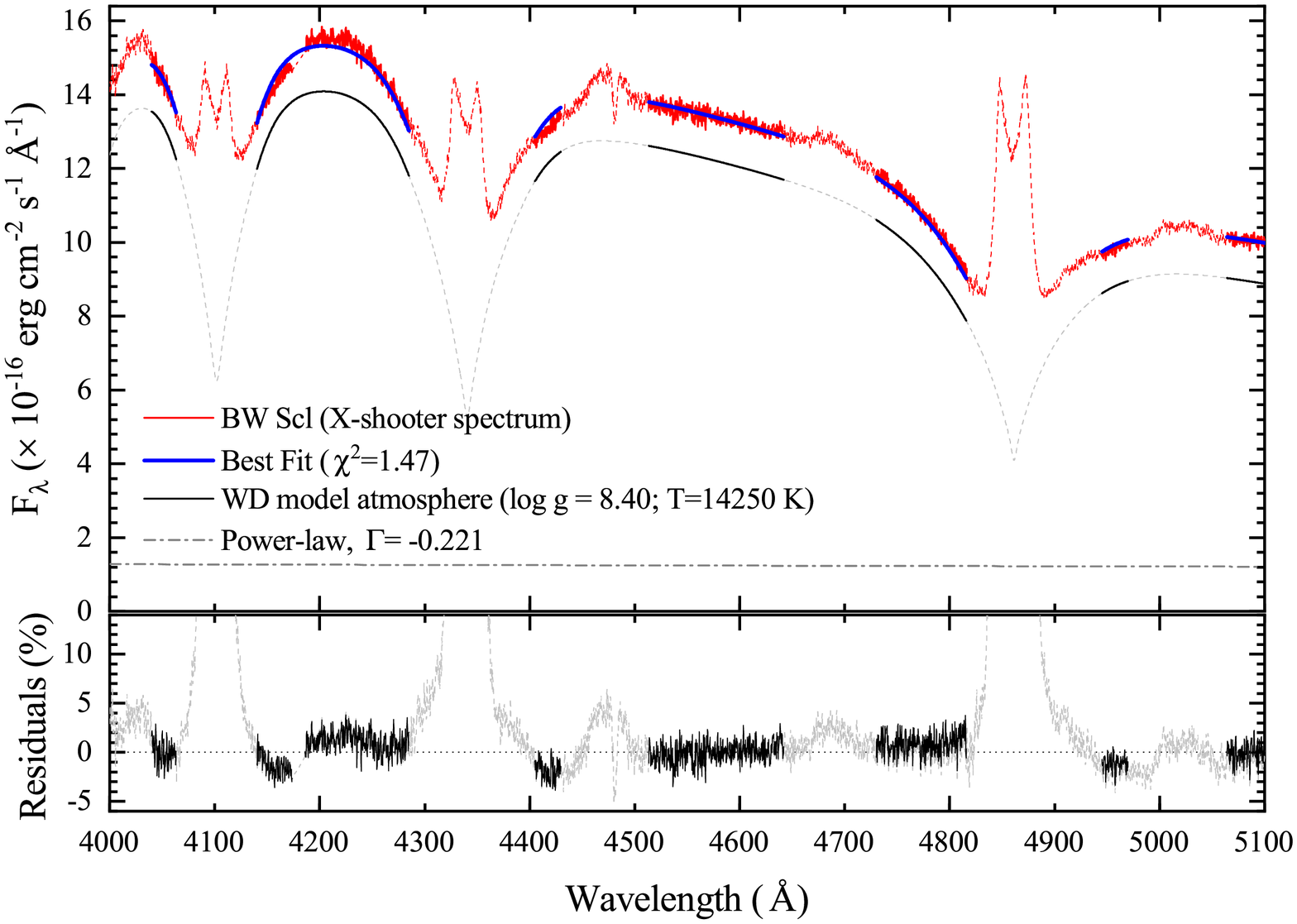} \includegraphics{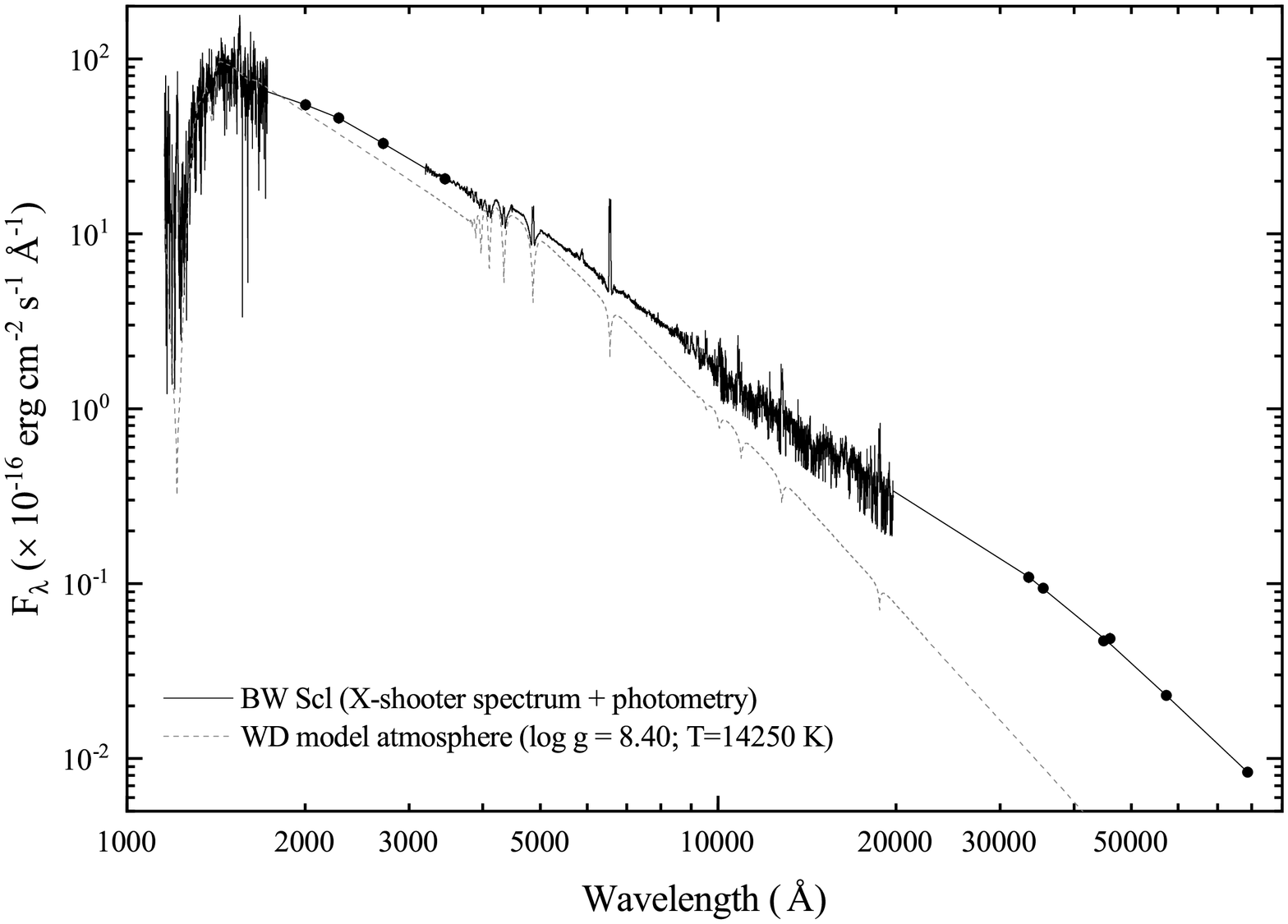}}
\caption{Left: the X-Shooter spectrum (red) along with the best-fitting model (blue).
The WD emission (black) and a power-law component (grey dash--dotted line) contribute 90 and 10 per cent of the total flux
at 4600 \AA, respectively. The masked spectral areas are plotted by dashed lines.
Right: the SED of BW~Scl, shown together with the best-fitting WD model spectrum.}
\label{Fig:SED}
\end{figure*}

Our fitting procedure can be described as follows. We use the flux calibrated average spectra, in which we consider only the
\Hbeta, \Hgamma\, and \Hdelta\ lines because the width of higher-order Balmer lines of the WD in BW~Scl becomes comparable with
that of the disc emission component. We cut out the emission cores of these lines and then perform the $\chi^2$ fit of the object
spectrum to a grid of synthetic spectra of DA WDs, to which the power-law flux was added.
We used a grid of \citet{Koester} model atmospheres\footnote{\citet{Koester} synthetic spectra of DA WDs were retrieved from
http://svo2.cab.inta-csic.es/theory/newov2/index.php?models=koester2} that cover a broad range of $T_{\rm eff}$ in steps of 250~K
and $\log g$ in steps of 0.25. Using linear interpolation from the grid, we also produced additional model spectra
for $\log g$ in step of 0.05.
The model spectra were convolved with
the appropriate Gaussian instrumental profile to match the spectral resolution of observed spectra. From an extensive testing of this technique
(\citealt{JussiThesis}; Neustroev \& Hedem\"{a}ki, in preparation\footnote{\url{https://vitaly.neustroev.net/research/wd-parameters/}}),
it has been shown that even for a large contribution of a power-law component ($\gtrsim$100\%), a low SNR$\sim$30, and relatively
broad emission cores eliminated from the fitting procedure (up to $\pm$4000 \kms), uncertainty estimates of parameters were within
our grid steps ($\Delta T_{\rm eff}$ = 250 K and $\Delta \log g$ = 0.25).\footnote{Obviously, these errors represent only the
ability of the model spectra to match the observed
spectra and most likely are underestimated. An example of more accurate analysis of the errors associated with measurements of
WD parameters can be found in \citet{Kepler07}. In partucular, using Monte Carlo simulations, they estimated an uncertainty of
around $\Delta T_{\rm eff}$ $\simeq$ 500 K and $\Delta \log g$ $\simeq$ 0.10 at SNR = 40 for the whole spectra fitting.}

The knowledge of the distance $d$ to the source helps to significantly increase the reliability of parameter estimates,
allowing the WD flux to be scaled correctly. The relation between observed fluxes $F_{\rm obs}$ and model Eddington fluxes
$H$ is
\begin{equation}
\label{Eq:FluxNorm}
    F_{\rm obs} = 4\pi H \, \frac{\Rwd^2}{d^2} \,,
\end{equation}
where \Rwd\ and $d$ are in the same units. In our fitting procedure, we link $\log g$ and \Rwd\ using an approximation formula
which is based on Nauenberg's mass--radius relation:

\begin{figure}
\resizebox{\hsize}{!}{\includegraphics{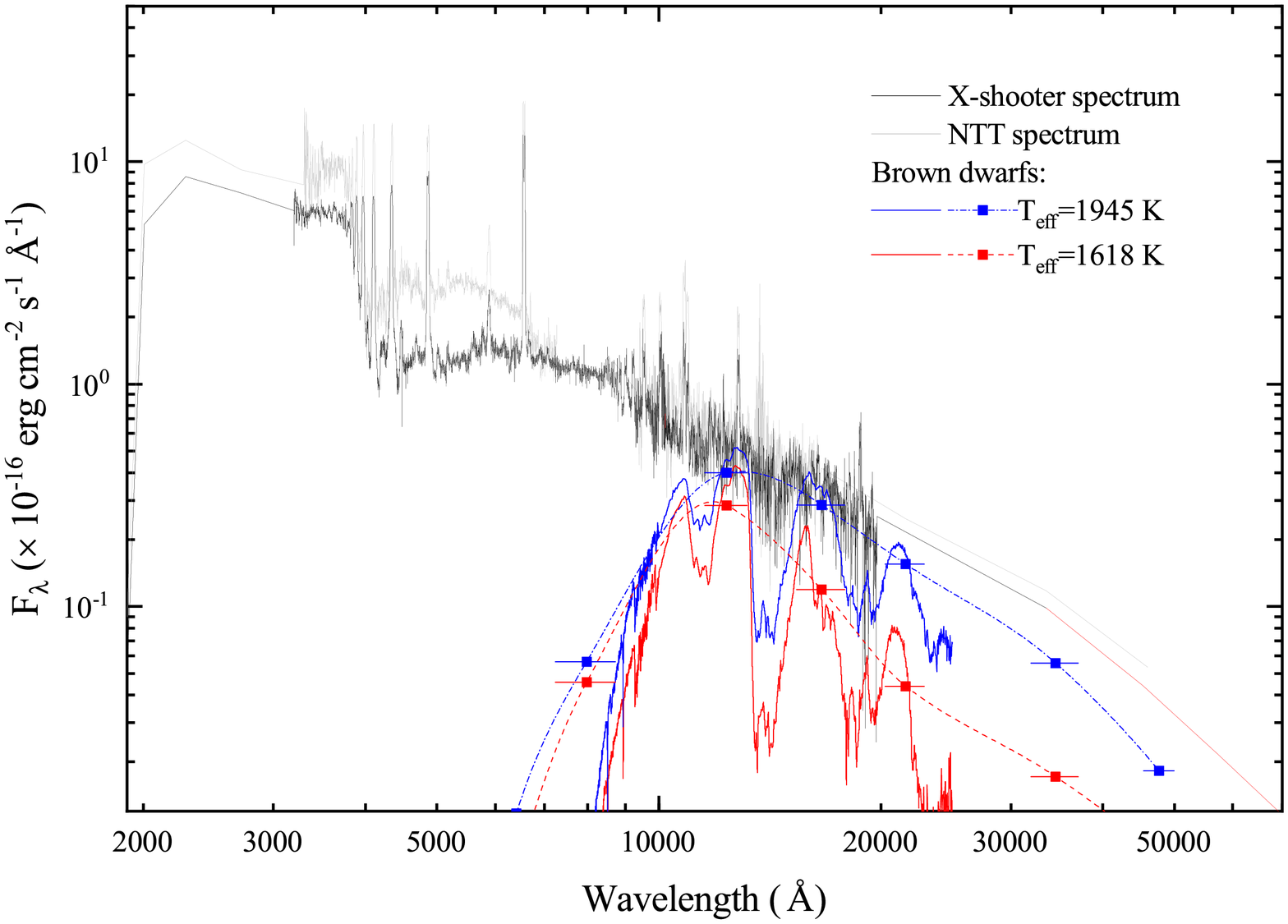}}
\caption{The accretion disc spectra, shown together with the theoretical SEDs and observed spectra of two brown dwarfs (see the 
text for details).}
\label{Fig:DiscSpec}
\end{figure}

\begin{equation}
\label{Eq:RwdLogg}
    \Rwd / R_{\sun} =(0.4074-0.0368*\log g)^2 \,.
\end{equation}
It is accurate to better than 1\% over the range 6.5 < $\log g$ < 9.35 (0.1~\Msun\ < \Mwd\ < 1.35~\Msun).

The surface gravity of the WD in BW~Scl is already known from the gravitational redshift, thus simplifying our problem
even further. This allowed us to fix $\log g$ at 8.40 and apply the described technique to each average spectrum of BW~Scl,
resulting in $T_{\rm eff}$=14250\,K for the X-shooter and NTT spectra\footnote{We point out that when $\log g$ is
allowed to vary, the best solution is also found at $\log g$  = 8.40, exactly consistent with the gravitational redshift.}
and 13250\,K for both the UVES spectra
(\autoref{Tab:WDparameters}). We note that the values of \Teff\ as derived from the UVES observations are 1000~K lower than that
from the X-shooter spectrum. Given that the UV and optical fluxes of BW~Scl were very stable for many years before the superoutburst
(see Section~\ref{Sec:LongTermPhot}), we assume (although cannot be sure) that this difference may not be real and that the UVES
\Teff's are underestimated due to low quality of those spectra. For this reason, we fixed \Teff\ at 14250~K and repeated the fitting
procedure for the UVES data.
The WD temperature found from the best quality data (X-shooter and NTT) is lower than that derived by \citet{Pala22}.
One of the reasons for that might be a smaller radius (due to a higher mass) of the WD in their solution.

\begin{table*}
\begin{center}
\caption{The WD temperatures and the disc contribution to the total flux at 4600~\AA. 
 Also shown the parameters of the Balmer emission lines in the disc spectra.}

\begin{tabular}{llcccccccccc}
\hline \hline
Data set    &$T_{\rm eff}$ (K) &$F_{\rm d}/F_{\rm total}$&\multicolumn{4}{c}{Equivalent width (\AA)} & \multicolumn{4}{c}{Flux (\tim{-14} \ergs)} & Balmer decrement \\
            &($\log g$=8.4)  &    ( \% )        & \Halpha&\Hbeta&\Hgamma&\Hdelta&\Halpha&\Hbeta&\Hgamma&\Hdelta&\Halpha\ : \Hbeta\ : \Hgamma\ : \Hdelta \\
\hline
UVES-1      &  13250         &      17          &    360 & 201 & 180 & 125 & 6.3 & 4.3 & 3.8 & 3.3 &  1.46 : 1.00 : 0.89 : 0.75  \\
            &  14250 (fixed) &       5          &    450 & 415 & 545 & 385 & 6.3 & 4.6 & 4.2 & 3.6 &  1.35 : 1.00 : 0.90 : 0.79  \\
UVES-2      &  13250         &      19          &    --- & 225 & 135 & 120 & --- & 4.8 & 3.5 & 3.4 &  \phantom{1.35} : 1.00 : 0.74 : 0.70  \\
            &  14250 (fixed) &       5          &    --- & 450 & 370 & 400 & --- & 4.9 & 3.8 & 3.7 &  \phantom{1.35} : 1.00 : 0.78 : 0.75  \\
X-shooter   &  14250         &      10          &    341 & 210 & 190 & 170 & 4.8 & 2.8 & 2.4 & 2.0 &  1.71 : 1.00 : 0.86 : 0.71  \\
NTT         &  14250         &      19          &    340 & 194 & 178 & 174 & 7.2 & 4.5 & 4.0 & 3.7 &  1.60 : 1.00 : 0.89 : 0.82  \\
\hline
\end{tabular}
\label{Tab:WDparameters}
\end{center}
\end{table*}

\begin{table*}
\begin{center}
\caption{The broad-band absolute magnitudes of the accretion disc derived from different data sets. Also shown the predicted absolute 
 magnitudes of brown dwarfs with $T_{\rm eff}$=1618 and 1945\,K and the radius 0.103\,$R_{\sun}$ \citep{KniggeCVevol}.}

\begin{tabular}{lcccccccc}
\hline \hline
Data set                      & $M_{U}$ & $M_{B}$ & $M_{V}$ & $M_{R}$ & $M_{I}$ & $M_{J}$ & $M_{H}$ & $M_{K}$ \\
\hline
Accretion disc: \\
UVES-1 ($T_{\rm WD}$=13250~K) &  12.17  &  13.21  &  13.19  &         &         &         &         &         \\
UVES-1 ($T_{\rm WD}$=14250~K) &  12.55  &  13.69  &  13.63  &         &         &         &         &         \\
X-shooter                     &  12.35  &  13.76  &  13.66  &  12.96  &  12.64  &  12.01  &  11.41  &  10.78  \\
NTT                           &  11.85  &  13.11  &  12.99  &  12.51  &  12.29  &  11.74  &  11.27  &  10.65  \\
\hline
Brown dwarf ($T_{\rm eff}$=1618\,K)         &  30.66  &  25.46  &  22.97  &  18.98  &  16.13  &  12.76  &  12.60  &  12.54  \\
Brown dwarf ($T_{\rm eff}$=1945\,K)         &  28.86  &  24.57  &  21.38  &  18.32  &  15.90  &  12.39  &  11.65  &  11.17  \\
\hline
\end{tabular}
\label{Tab:ADparameters}
\end{center}
\end{table*}

\subsection{Spectral energy distribution and accretion disc spectra}
\label{Sec:SED}

Combining our multiwavelength data with the archival UV and NIR observations, we reconstructed the spectral energy
distribution (SED) of BW~Scl in the UV-optical-NIR wavelengths at the time of the X-shooter (\autoref{Fig:SED}) and
NTT observations.
As seen in \autoref{Fig:SED} and \autoref{Tab:WDparameters}, the UV-optical-NIR spectrum is dominated by the WD component
which produces around 80-90 per cent of the flux in the blue wavelength range. However, a non-WD contribution increases
with longer wavelengths and becomes dominant at about 13000~\AA. The SED appears very smooth. No sign of the donor
star is visible, but one can notice a hint of a ``knee'' at $\sim$30000--40000~\AA. Here we have to note that
although in short-period CVs with a main-sequence donor star the NIR hump in the SED should not be expected to be as
pronounced as in longer period CVs with a low mass-transfer rate (as seen, e.g., in RZ~Leo -- \citealt{NeustroevSSS}),
it can still be detectable \citep[see figure~1 in][]{NeustroevKniggeZharikov}. On the other hand, a sub-stellar donor
in a period bouncer is expected to be so dim that the WD and even relatively weak accretion disc can outshine it.

In order to put an upper limit on the donor contribution to the total system light, we subtracted the underlying WD spectrum
from the observed SEDs. This allowed recovering not only a non-WD continuum but also higher-order Balmer emission lines
which were sitting inside the WD absorption troughs. In the following, we call the resulting spectra as the spectra of the
accretion disc. We admit that they are contaminated by the donor star, but we show below that the contribution of the donor
is very low.

\autoref{Fig:DiscSpec} shows the disc spectra recovered from the X-shooter and NTT data sets. The Balmer and Paschen continuum
flux was larger during the post-outburst NTT observations. 
The spectra display a strong Balmer jump and a notable Paschen jump, both in emission. Overall, the spectra strongly
resemble those produced in a hydrogen gas slab \citep[compare with, e.g. fig. 12 in][]{Pala-QZLib}. While the disc spectra
will be analysed in detail elsewhere, here in Tables~\ref{Tab:WDparameters} and \ref{Tab:ADparameters} we outline different
parameters of the continuum light and of the most prominent Balmer lines that were measured from the accretion disc spectra.
In particular, the spectra were convolved with the standard Johnson and Kron--Cousins $UBVR_cI_cJHK$ filter bandpasses, and
the broad-band absolute magnitudes were determined. Note that although the magnitudes vary slightly from spectrum to spectrum,
$M_{V}$ appears consistent with an indirect estimate presented by \citet[table~3]{PattersonDist}.

We can now compare the observed disc spectra with the SEDs of stars of different temperatures. For this, we used tables 2, 5,
and 6 from \citet{KniggeCVevol} which provide the absolute magnitudes of the donor along the CV evolution sequence. We extracted
the magnitudes of a few stars with $T_{\rm eff}$ between 1320 and 2254\,K, and converted them into fluxes, scaling to the
distance of BW~Scl and the radius of its donor (0.103\,$R_{\sun}$). We also used observed brown dwarf spectra, which were scaled
to be in accordance with the above-predicted fluxes. These spectra exhibit a characteristic series of broad peaks near 1.08, 1.27,
1.65, and 2.08 $\mu$m. Taking into account the noise level in the NIR spectra, we put a conservative upper limit on the donor
temperature to be $T_{\rm eff,2}\lesssim$1\,600~K, otherwise the mentioned flux peaks will become obvious \emph{even assuming} that
the disc spectrum declines strongly toward longer wavelength in the NIR.
However, the donor temperature should not be much lower than that because we see a sign of the donor
\KI\ absorption lines (12432 and 12522~\AA) in the averaged spectrum corrected for orbital motion of the donor star (see the inset
panel in \autoref{Fig:Xsh-spec}). The spectra and calculated SEDs of two brown dwarfs with $T_{\rm eff}$=1618 and 1948\,K are
shown in \autoref{Fig:DiscSpec} together with the disc spectra. We find it instructive to include their magnitudes into
\autoref{Tab:ADparameters}, allowing for a direct comparison between the donor and observed fluxes.

By integrating the disc SEDs over all wavelengths, we can put a conservative upper limit on the bolometric luminosity of the
accretion disc and on the mass-accretion rate. The latter can be calculated using
\begin{equation}
 \dot M_{\rm acc} = \frac{2 L_{\rm d} R_{\rm wd}}{G M_{w\rm d}} \,.
\end{equation}
We find these parameters for the X-shooter spectrum to be $L_{\rm d}\lesssim$3.2\tim{30} \ergpers\ and $\dot M_{\rm acc}\lesssim$
3.7\tim{13} g s$^{-1}$ = 5.9\tim{-13} \Msun\ yr$^{-1}$, and $L_{\rm d}\lesssim$4.0 \tim{30} \ergpers\ and $\dot M_{\rm acc}\lesssim$
= 4.6\tim{13} g s$^{-1}$ = 7.4\tim{-13} \Msun\ yr$^{-1}$ for the NTT spectrum. These estimates include the
contribution from the donor which is less than 8 per cent.
We also point out that the disc luminosity $L_{\rm d}$ appears to be a few times larger than the X-ray luminosity (see
Section~\ref{Sec:XrayData}). This is consistent with the observations of other short-period CVs \citep{NeustroevSSSX,Amantayeva21},
and with the disc instability model according to which the mass-accretion rate in the quiescent disc is expected to decrease
steeply with decreasing radius \citep{Cannizzo93,Ludwig94}. The latter means that accretion rates at the WD surface should be
smaller than in the outer disc. Thus, the X-ray flux which almost certainly originates in regions very close to the WD surface
should indeed be quite low.

\begin{figure*}
    \resizebox{\hsize}{!}{\includegraphics[angle=0]{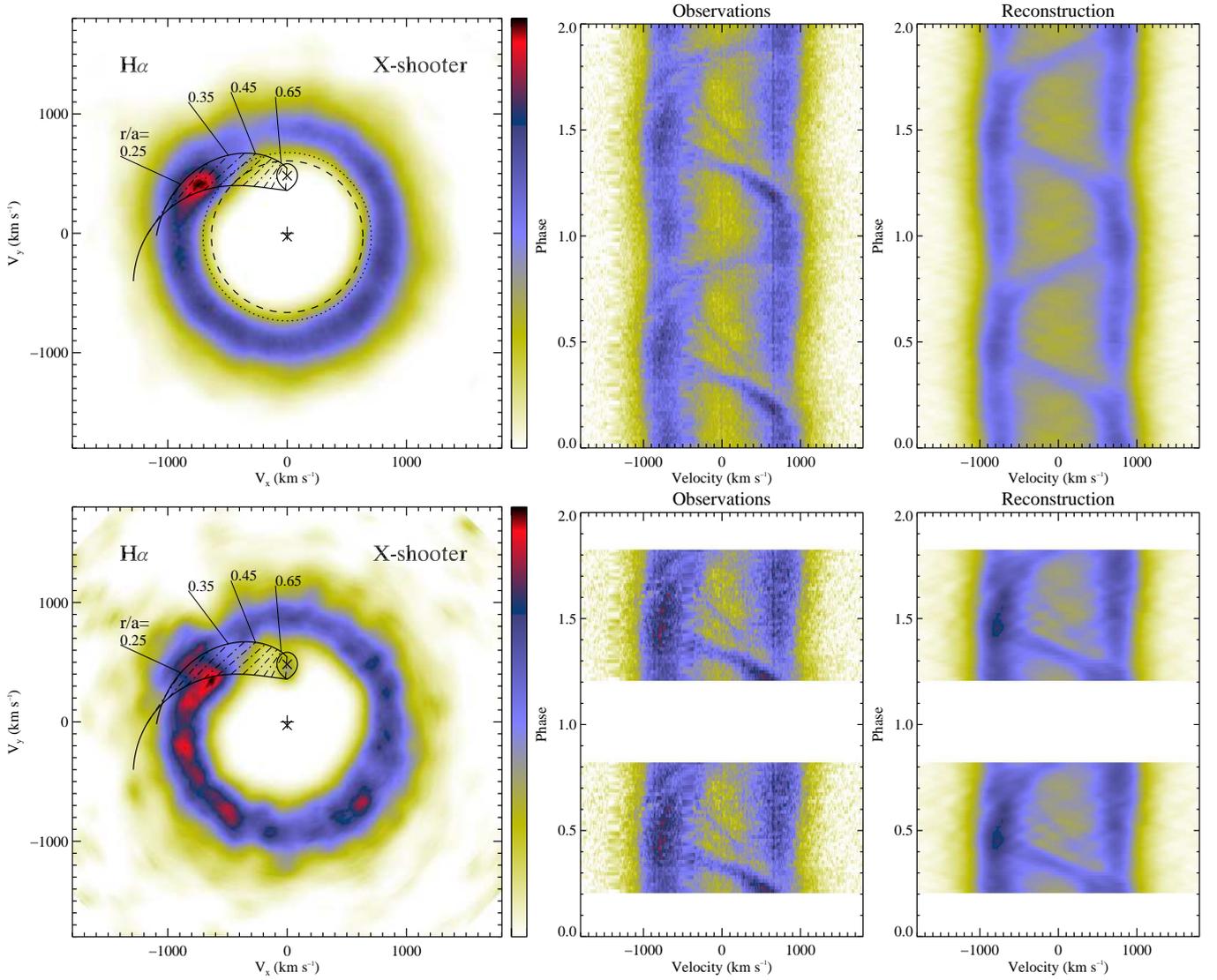}}
    \caption{The Doppler maps and corresponding observed and reconstructed trailed spectra of the \Halpha\
             emission line. The upper map is calculated using the whole set of spectra, whereas for the lower
             map only 60 per cent of spectra between phases 0.2--0.8 were used.
             Marked on the maps are the position of the WD (lower cross), the centre of
             mass of the binary (middle cross) and the Roche lobe of the donor star (upper bubble
             with the cross). The dashed and dotted circles show the tidal truncation radius $r_{\rm max}$ and
             the 3:1 resonance radius $r_{\rm 3:1}$, respectively.
             The dashed lines connect the velocity of the ballistic gas stream (lower curve) and the velocity
             on the Keplerian disc along the gas stream (upper curve) for the same points at distances labelled
             along the upper curve (in $r$/$a$ units). These lines are separated by 0.05$r$/$a$.
             }
    \label{Fig:DopMapHa}
\end{figure*}

\subsection{Doppler tomography}
\label{Sec:DopMap}

\begin{figure*}
    \resizebox{\hsize}{!}{\includegraphics[angle=0]{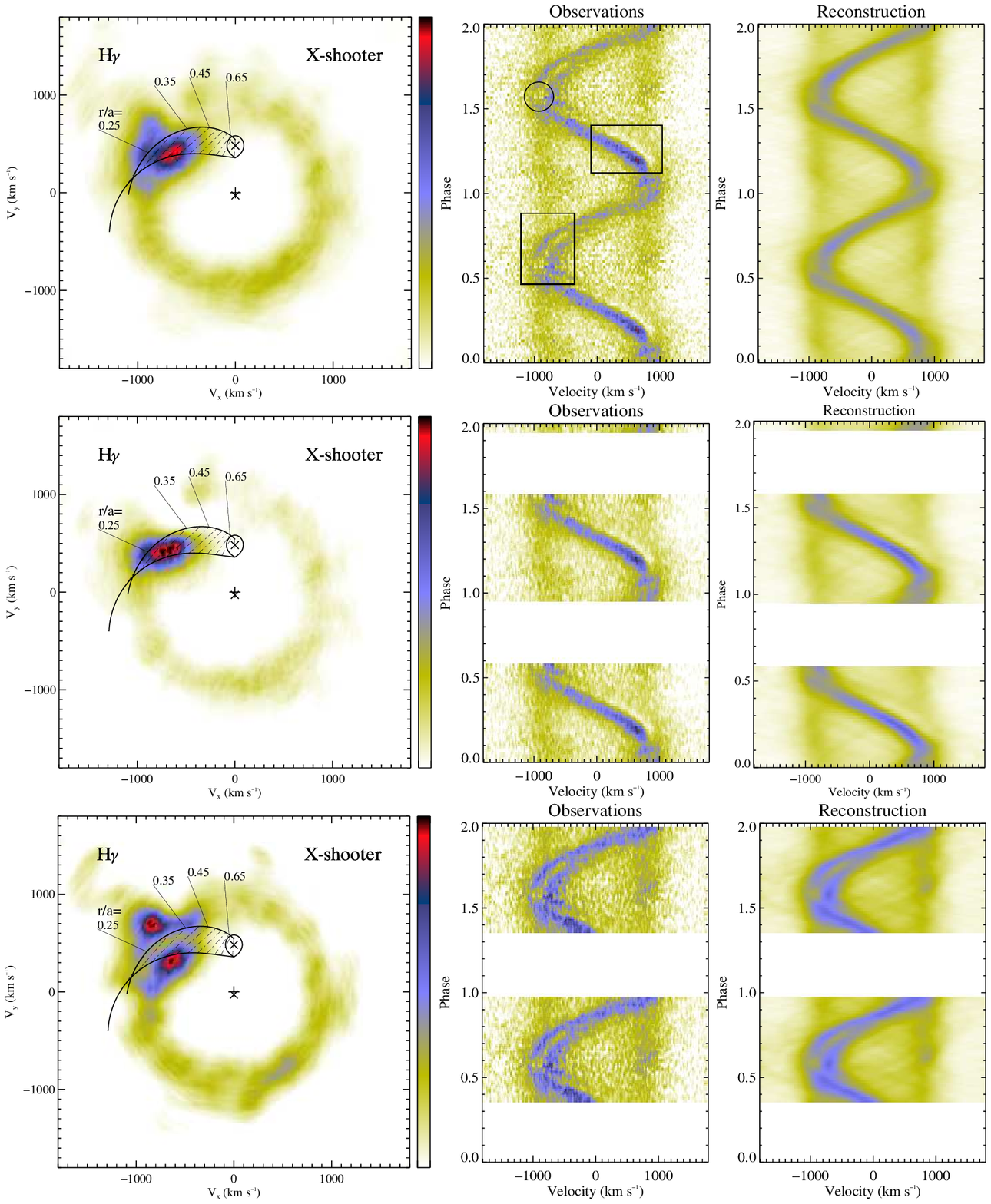}}
    \caption{Same as \autoref{Fig:DopMapHa}, for \Hgamma.
    Marked on the observed trailed spectrum are the positions of the S-wave split and the S-wave shadow (the left and right
    rectangles, respectively), and the S-wave disappearance around phase 0.6 (circle).
    The middle and lower maps were calculated using 60 per cent of spectra centred on phases 0.25 and 0.65, respectively.}
    \label{Fig:DopMapHg}
\end{figure*}

\begin{figure*}
    \resizebox{\hsize}{!}{\includegraphics[angle=0]{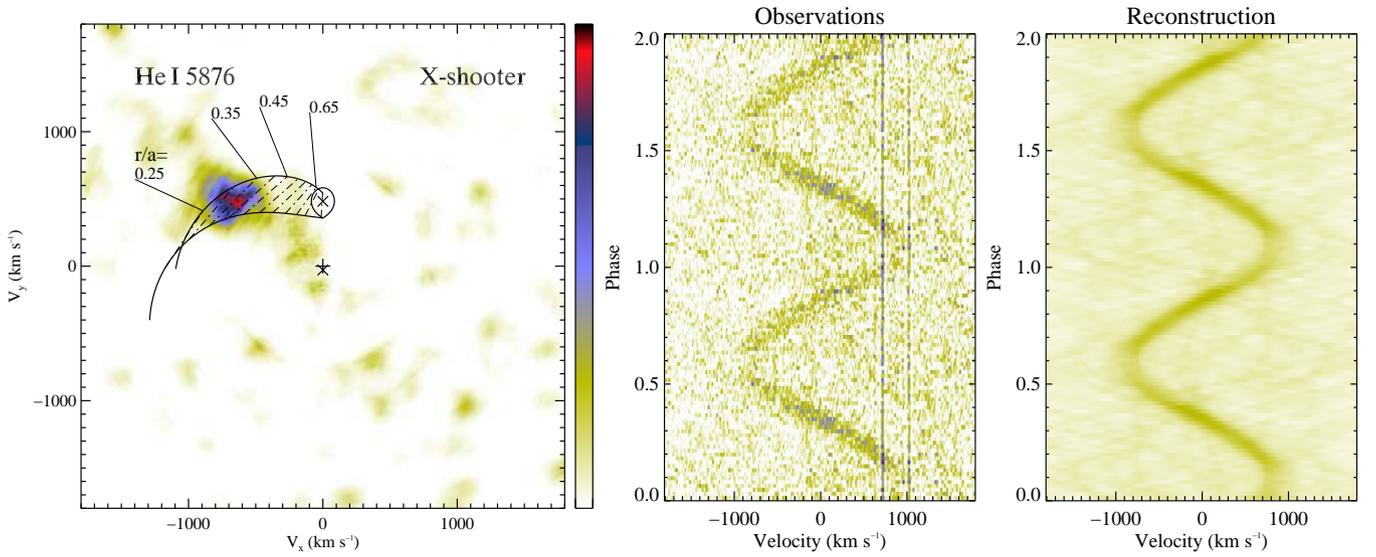}}
    \caption{Same as \autoref{Fig:DopMapHa}, for \HeI\ 5876~\AA. In the observed trailed spectrum, the Na\,D night-sky lines are seen.
             }
    \label{Fig:DopMapHe5876}
\end{figure*}

\begin{figure*}
    \resizebox{\hsize}{!}{\includegraphics[angle=0]{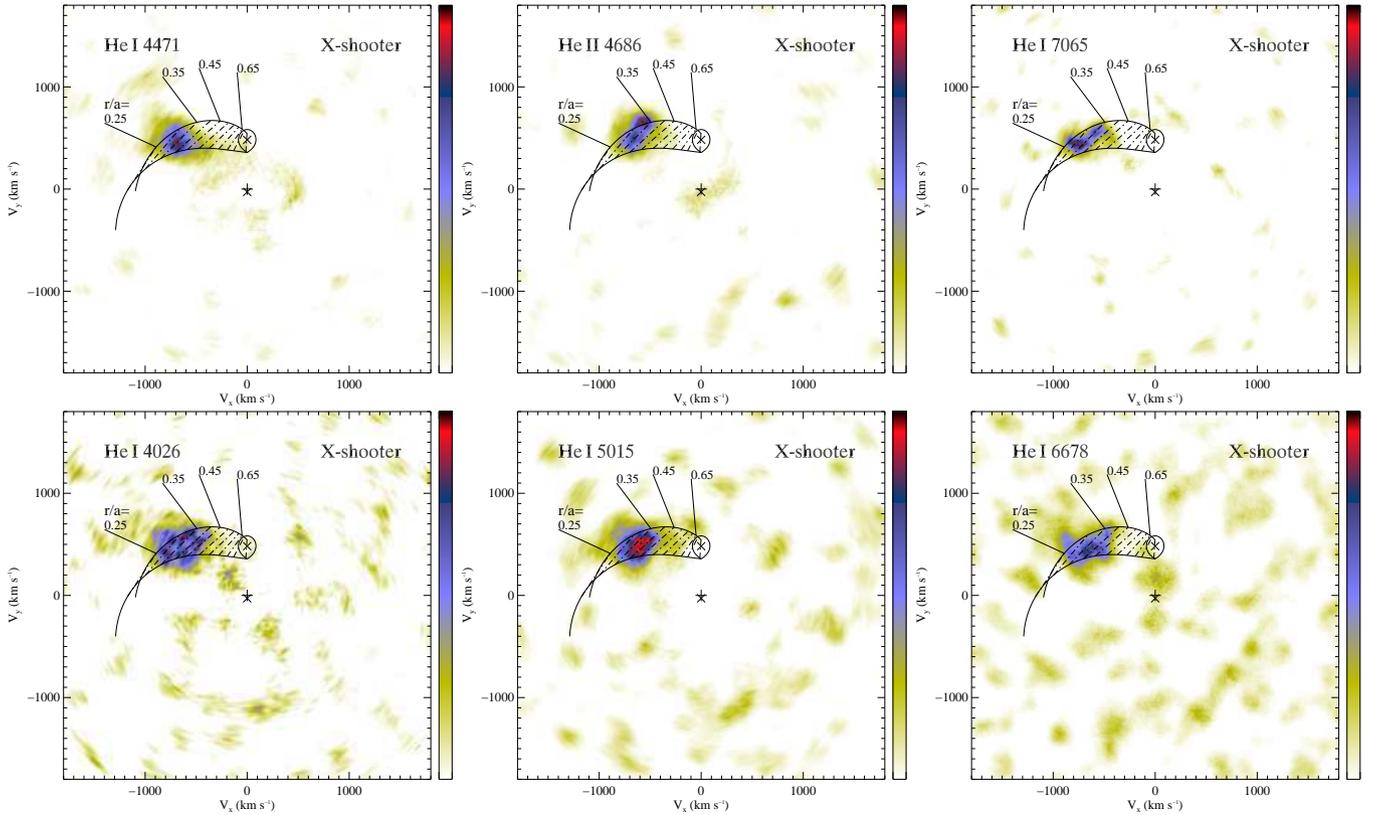}}
    \caption{The Doppler maps of selected \HeI\ and \HeII\ lines.             }
    \label{Fig:DopMapHe}
\end{figure*}

The trailed spectra of BW~Scl exhibit a mixture of different variable emission components. In order to
study the sources of emission in more detail, we used Doppler tomography \citep{MarshHorne88, Marsh2001}.
We produced a large number of tomograms, using the code developed by \citet{Spruit}. Bearing in mind the
unprecedented quality of the X-shooter data, in this section we show the most representative lines from
this data set only. Figures~\ref{Fig:DopMapHa}, \ref{Fig:DopMapHg}, and \ref{Fig:DopMapHe5876} display
the Doppler maps of \Halpha,
\Hgamma, and \HeI\ 5876 together with the trailed spectra and their corresponding reconstructed counterparts.
However, supplementary Figures~\ref{FigApp:DopMapHaAll}, \ref{FigApp:DopMapHbAll}, and \ref{Fig:DopMapHb}
provide a comparison of Doppler maps of \Halpha\ and \Hbeta\ from different data sets. They look
very consistent to each other, indicating no notable difference in the disc structure at different epochs.
Also, as seen from \autoref{Fig:DopMapHe5876}, the helium lines represent a simple case in terms of
their orbital variability, so in \autoref{Fig:DopMapHe} we only show the maps of the He lines, omitting
unnecessary trailed spectra.

We start the discussion of Doppler tomography with the \Halpha\ line as it is the strongest line in the spectra
(see the upper row of \autoref{Fig:DopMapHa}, and also \autoref{FigApp:DopMapHaAll}).
The maps are dominated by a ring of disc emission and a compact, relatively weak hotspot emission which is
located in the fourth quadrant ($-V_{x}$,$+V_{y}$) of the maps. We note that all Balmer lines display a gap
in the upper part of the disc ring. In addition to different marks on all the maps
that facilitate interpreting the tomograms (see the captions to \autoref{Fig:DopMapHa}), for \Halpha\ we also
show the circles representing velocities at the tidal truncation radius $r_{\rm max}$, $\upsilon_{\rm min}$,
and at the 3:1 resonance radius $r_{\rm 3:1}$, assuming a circular Keplerian flow in the disc. It is seen
that the emission extends beyond $r_{\rm 3:1}$ and is close to the truncated orbit. This is in accordance with
other studies reporting that the accretion disc in CVs is always extended to its truncation limit
\citep{NeustroevHT,NeustroevHT2,Amantayeva21}.

\begin{figure*}
\resizebox{0.74\hsize}{!}{
\includegraphics{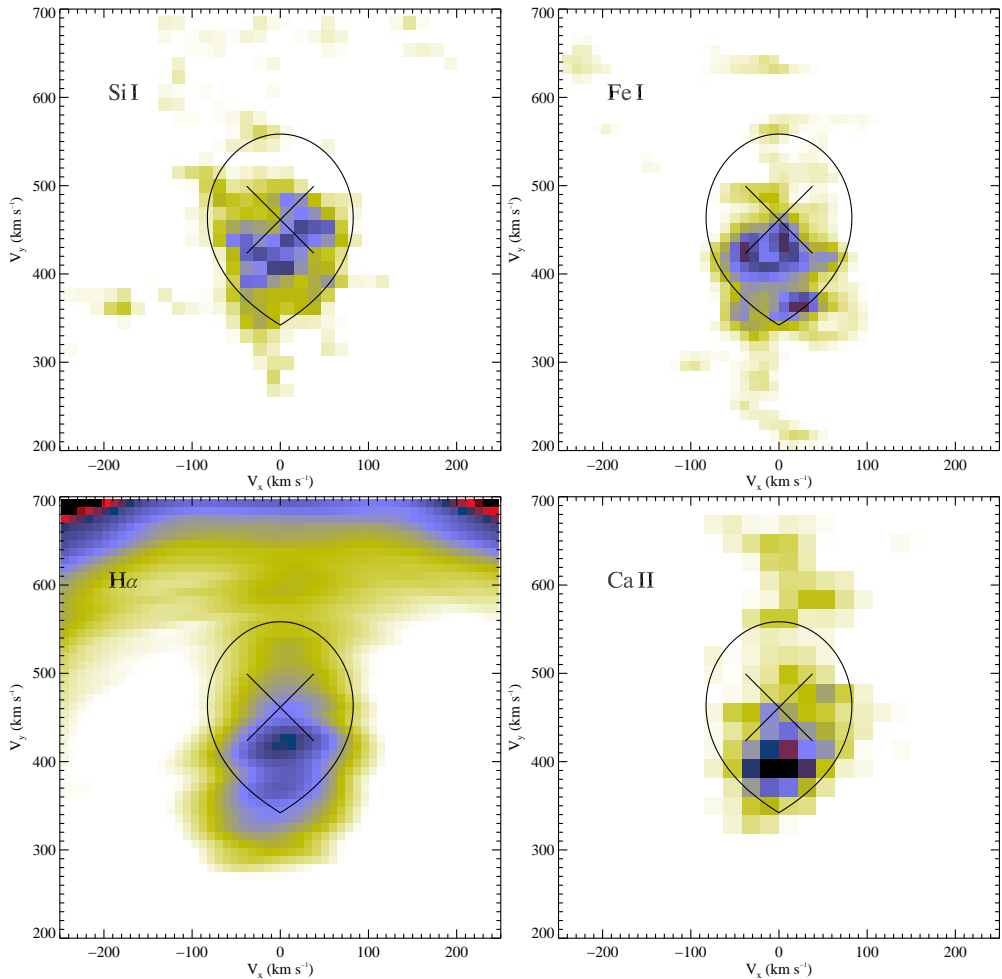}
}
\caption{Zoomed parts of the tomograms of the lines in which the donor star is detected (\SiI, \FeI, \Halpha, and \CaII).
The maps are centred around the donor area.}
\label{Fig:DonorDopMap}
\end{figure*}

\begin{figure*}
    \resizebox{\hsize}{!}{\includegraphics[angle=0]{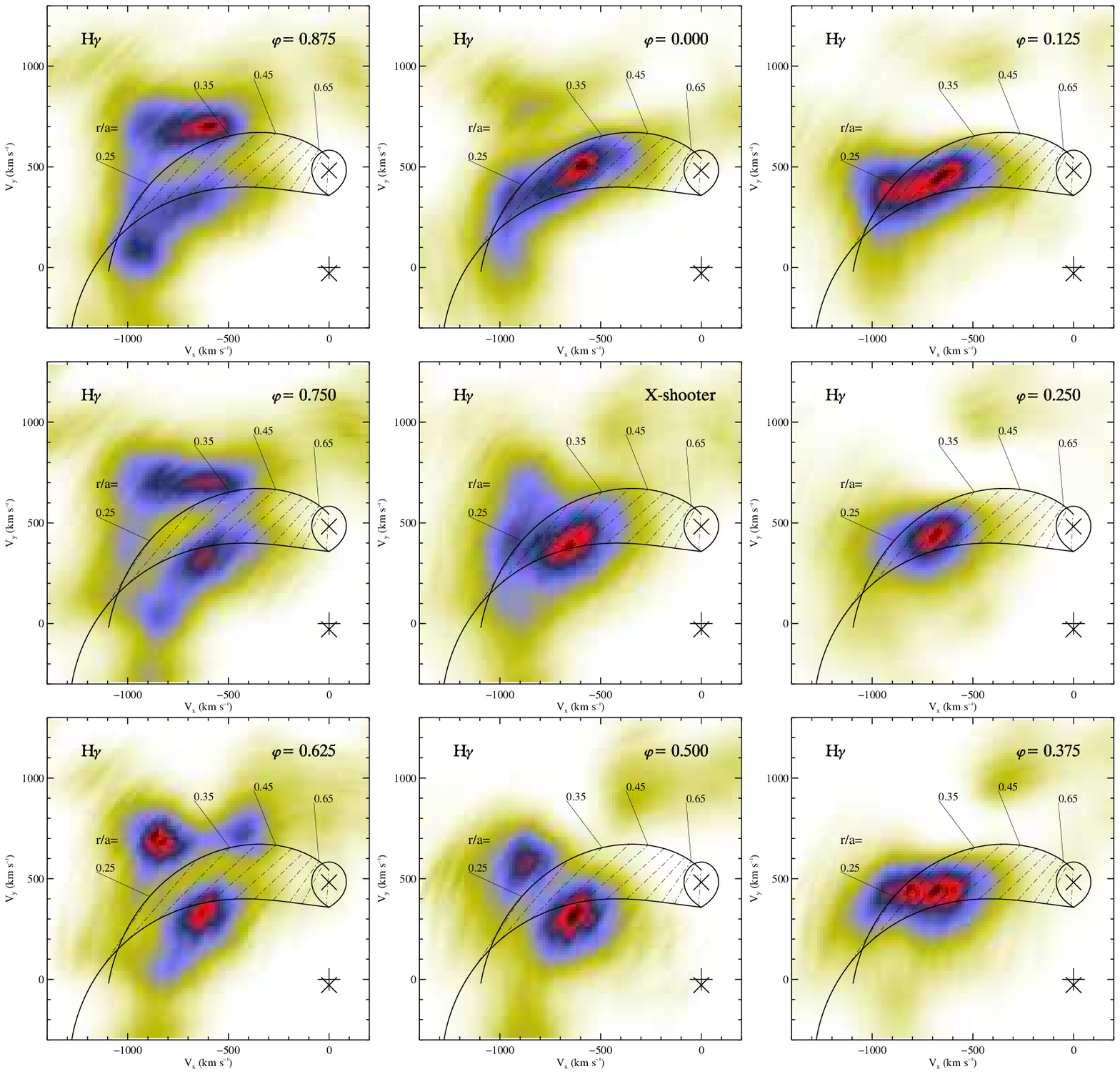}}
    \caption{The evolution of the appearance of the hotspot area on Doppler maps of \Hgamma, which were calculated using 40 per cent
             of spectra centred on phases shown on the maps. The central map was calculated using all the spectra. The maps are scaled
             according to their minimum and maximum values.
             }
    \label{Fig:DopMapDynHS}
\end{figure*}

The hotspot emits at a very wide range of velocities. It occupies a larger than 500$\times$400 \kms\ area on the Doppler
maps, reflecting a large width of the hotspot S-wave. The latter can be compared with e.g. the donor S-wave in \Halpha,
or the Na\,D night-sky lines, which are seen in the trailed spectrum of \HeI\ 5876 in \autoref{Fig:DopMapHe5876}.
The location of the hotspot area on the maps is basically consistent with the trajectory of the gas stream, showing
a mixture of the ballistic and Keplerian velocities along the stream. Although the hotspot has a complex, arrow-like structure,
the peak of spot emission agrees better with the ballistic stream trajectory and locates well inside the disc, as close
to the WD as $r_{\rm hs}$$\approx$0.35--0.40\,$a$$\approx$0.6--0.7\,$r_{\rm d}$.

The \Halpha\ tomogram shows only a weak sign of the donor star. Comparing the observed trailed spectrum with its
reconstructed counterpart (the central and right plots in the upper row of \autoref{Fig:DopMapHa}), one can notice
that the latter is also missing the sharp emission component produced by the donor. The reason for it is that this
weak feature is visible only during half of the orbital period and disappears during another half. This violates an
assumption of Doppler tomography that all points are equally visible at all times \citep{Marsh2001}. In order to
overtake this problem, we calculated another tomogram using only about 60 per cent of spectra between phases 0.2--0.8.
Now the donor is obvious in both the map and the corresponded
reconstructed trailed spectrum (the lower row of \autoref{Fig:DopMapHa}). By using the same approach, we also
calculated the maps of other lines (\SiI, \FeI, and \CaII) in which we detected the donor star. Zoomed parts of
these maps that are centred around the donor area are shown in \autoref{Fig:DonorDopMap}.

It is interesting that the hotspot looks differently on the \emph{donor-on} map of \Halpha\ compared with the tomogram,
calculated using the whole phase range of spectra. This may indicate that the radiation emitted by the hotspot is
anisotropic. To study the hotspot area and the presence of anisotropy in more detail, we used higher order Balmer lines
in which the hotspot reaches a higher contrast with respect to the disc than in \Halpha\ (see the maps and trailed
spectra of \Hgamma\ and \Hbeta\ in Figures~\ref{Fig:DopMapHg} and \ref{Fig:DopMapHb}, respectively).
We point out that, similar to the case of the donor emission, the trailed spectra reconstructed from the maps of the whole phase
range are missing some distinctive features of the S-wave which are clearly visible in the observed trailed spectra. The most
prominent of them are the S-wave splitting and shadow, marked on the observed trailed spectrum in Figure~\ref{Fig:DopMapHg}
(see also Figure~\ref{Fig:ProfileFeatures}).  It is likely that these
two features are related to each other as the shadow is seemingly transitioning into one of the split S-waves at phase $\sim$0.5,
or is creating this apparent split (see the trailed spectra in Figures \ref{Fig:DopMapHg} and \ref{Fig:DopMapHb}). Another
interesting feature has first been mentioned by \citet{SpruitRutten} in their study of WZ~Sge. They detected the disappearance
of the S-wave around phase 0.6, which we also confirm for the bluer, higher velocity S-wave of BW~Scl. Some of these features
in WZ~Sge were interpreted by \citet{SpruitRutten} in terms of finite optical-depth effects.

Whatever is the reason for such a behaviour of the S-wave, its source violates another assumption of Doppler tomography:
it changes visibility or projected size, resulting in a changed flux. To ease this problem, we applied the approach used
above to recover the donor emission. We calculated 2 Doppler maps for each of \Hbeta\ and \Hgamma, using again 60 per cent
of spectra centred on phases 0.25 (0.95--0.55) and 0.65 (0.35--0.95). The new reconstructed trailed spectra reproduce the
observed ones in great detail. Both the S-wave splitting, the shadow, and even the disappearance of the S-wave at phase 0.6
are clearly visible now. Amazingly, the visual appearance of the hotspot area on the new maps has changed dramatically. The
hotspot looks absolutely different when watching from different directions, confirming the strong anisotropy of the emission.

The spectra centred on phase 0.25 produce on the Doppler map a long and relatively narrow stream of emission starting at
$r$$\approx$0.6\,$a$ from the WD, at the disc edge, and basically following the ballistic trajectory until $r$$\approx$0.25\,$a$.
Above this emission region, there is a horizontally extended area with no emission from the disc. It is this feature that causes
the S-wave shadow. In the following, we call it as `an empty spot'. Its position on the map ($V_{x}$$\approx$$-$400,
$V_{y}$$\approx$+800) roughly coincides with the gap in the disc, which we mentioned above.

Both these features appear inverted on the second map which is based on the spectra centered on phase 0.65. The former stream
of emission has now a brightness of the underlying disc, but on both sides of it, above and below on the map, it is accompanying
by two newly appeared emission streams. The lower stream roughly follows the ballistic trajectory, but located slightly below it
on the map. The upper emission stream follows Keplerian velocities along the trajectory of the gas stream, although also located
above it. The position of the latter emission feature on the map roughly coincides with the empty spot.

Comparing these two maps, we can also notice a different contrast of the hotspot with respect to the underlying disc. The maps
are scaled according to their minimum and maximum values. The weaker appearance of the disc on the maps centred on phase 0.25
means that from this direction the hotspot appears brighter and more compact than from the perpendicular direction.

A sharp contrast between the two maps motivated us to calculate a sort of a dynamical Doppler map of \Hgamma\ which allow examining
the evolution of the appearance of the hotspot area.\footnote{The animation may be viewed at \\
\url{https://vitaly.neustroev.net/researchfiles/bwscl/}}
To this end, we used shorter subsets of spectra\footnote{For this map, we used subsets of spectra consisting of 40 per cent of the whole X-shooter
data set.
}
whose midphases vary from 0 to 1. A static version of the map in which the individual tomograms are centred around the hotspot area
and scaled according to their minimum and maximum values is shown in \autoref{Fig:DopMapDynHS}. The full velocity range map scaled
according to the average disc brightness can be found in \autoref{Fig:DopMapDynDisc}.

The trailed spectra of other than hydrogen lines do not show the S-wave splitting and shadow, although the disappearance of the
S-wave at phase 0.6 can be detected in the strongest \HeI\ lines (\autoref{Fig:DopMapHe5876}). As a result, the Doppler maps of
these lines indicate at most only a weak presence of anisotropy of the hotspot emission. The hotspot area in the helium lines is
much more compact than in the hydrogen lines (\autoref{Fig:DopMapHe}), although the maximum of the helium emission is also located
well inside the disc, at the same range of distances from the WD $r_{\rm hs}$$\approx$0.35--0.40\,$a$. Similarly, a weak helium
emission can also be traced along the stream until the disc edge at $\sim$0.6\,$a$. The \HeII\ 4686~\AA\ line is very similar in
appearance to the \HeI\ lines.

\begin{figure*}
\resizebox{\hsize}{!}{
\includegraphics{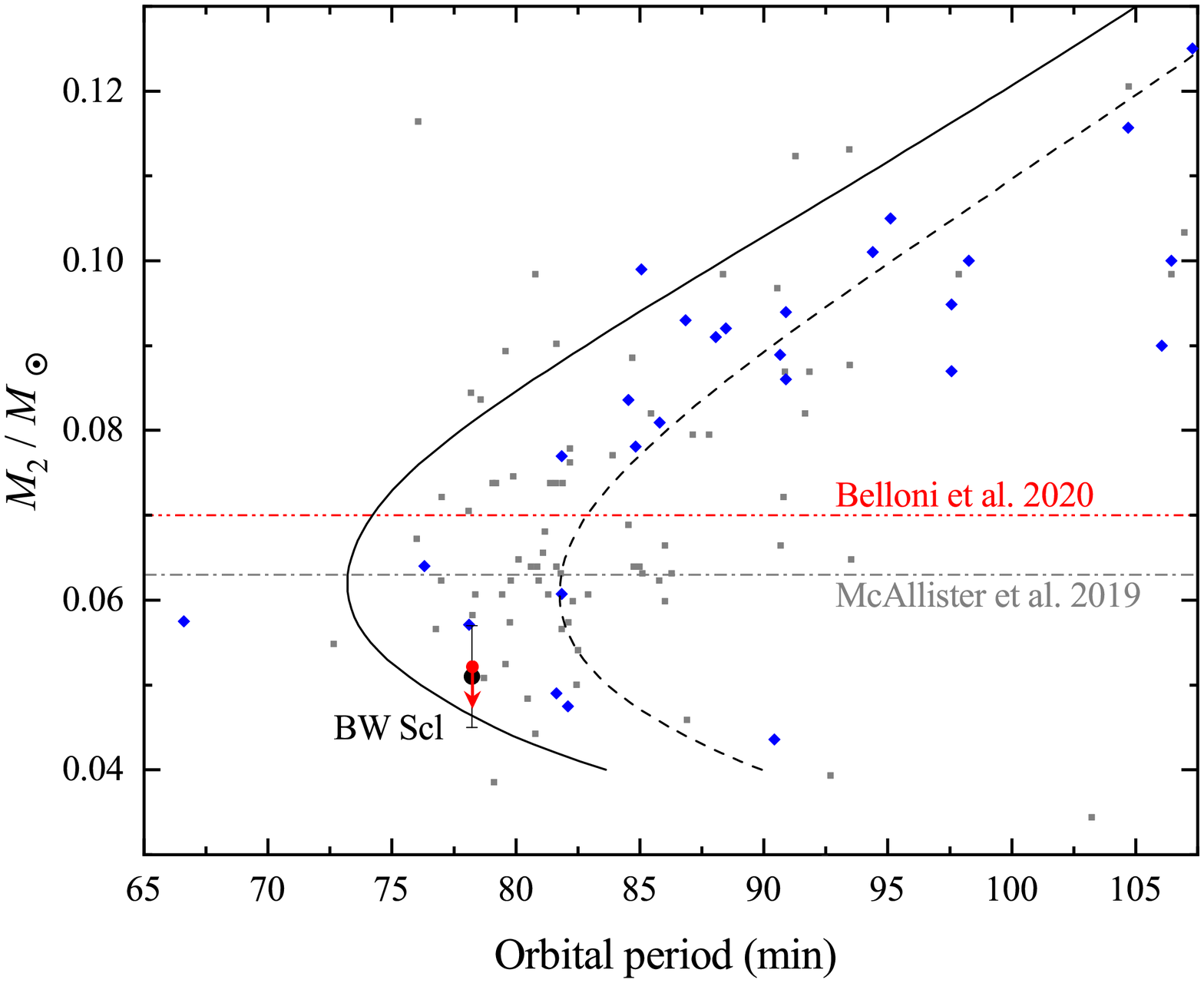}\includegraphics{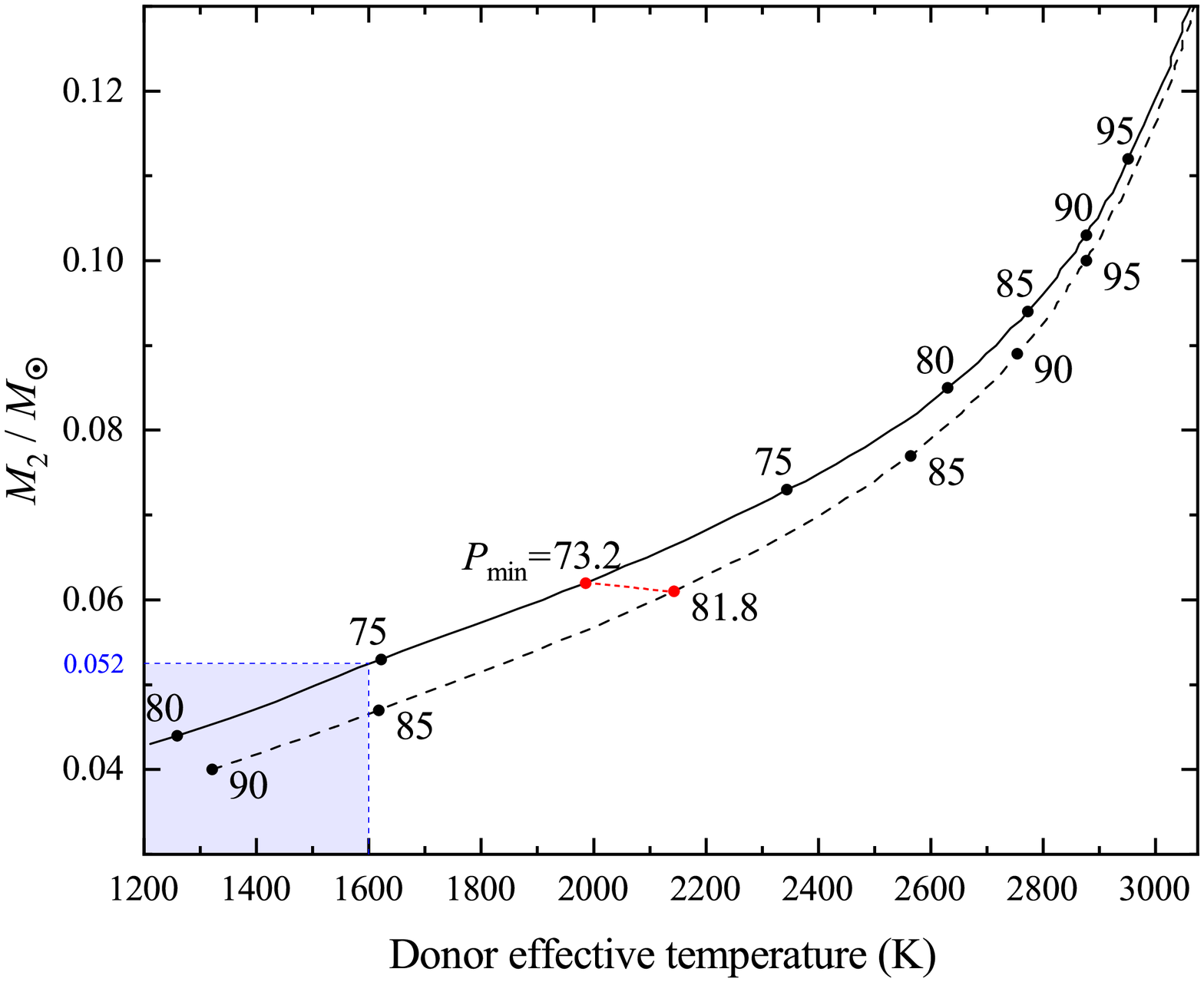}
}
\caption{Left: donor masses in short-period CVs versus orbital periods. The large black dot with error bars represents
a dynamically measured donor mass in BW~Scl, whereas the red dot with an arrow shows the upper limit of the mass estimated
from the donor temperature. The blue and grey points represent masses determined by eclipse modelling from \citet{McAllister19}
and the stage A superhump method \citep{Kato22}, for which error bars have been omitted for clarity. The two horizontal lines
indicate the donor mass at which the period bounce occurs, according to \citet{McAllister19} and \citet{Belloni20}.
Right: donor masses versus donor temperatures in short-period CVs. The black solid and dashed lines represent the `optimal' and
‘standard' evolutionary tracks from \citet{KniggeCVevol}, respectively; the red dotted line connects $P_{\rm min}$ in these tracks.
The dashed blue lines and shaded region show the range of masses for a donor with $T_{\rm eff,2}\leq$1600~K.
}
\label{Fig:EvolTracks}
\end{figure*}

\section{Discussion}
\subsection{Evolutionary status of BW Scl}

An accurate characterization of donor stars in short-period CVs (in particular, the measurement of their mass and effective
temperature/spectral type) is a difficult task. This requires the combined use of different spectral lines that originated in
the donor photosphere. These lines are usually very weak in the spectra of short-period CVs and can hardly be detected. The
problem becomes even more complex in the case of period-bouncers. The spectra of low-temperature donors in these systems,
brown dwarf-like objects, peak in the NIR where obtaining high-quality spectra is challenging. As a result, only very few
period-bounce candidates in quiescence revealed their donors in spectra (\citealt{HarrisonWZ,HarrisonWZ2,Pala-SDSS1238},
see also \citealt{Littlefair03} for discussion).

Dynamical information about the donor star can potentially be derived through the detection of irradiation-induced
emission lines. Although such lines are usually detected in the spectra of longer period CVs \citep{Harlaftis99, Schwope2000},
sometimes they are also seen in short-period dwarf novae during their superoutbursts and/or the following decline
\citep{SteeghsWZ}. The latter becomes possible because of significant compressional heating of the WD during outbursts
up to a few$\times$10\,000~K \citep[see e.g.][]{Long03,Bullock11}. It is, however, generally accepted that after subsequent
cooling and in proper quiescence the WDs in short-period CVs are not hot enough
\citep[the observed $T_{\rm eff}$$\lesssim$15\,000~K;][]{Pala17,Pala22} to excite, at detectable levels, emission lines
from the inner hemisphere of the donor star. Indeed, to the best of our knowledge, no such detections were reported to date.

Nevertheless, this study demonstrates that, at least under certain conditions, the irradiation-driven lines can
be strong enough to be seen. These lines allowed us to make a good estimate for the radial velocity amplitude of the donor
star in BW~Scl and to derive accurate system parameters. The donor is found to have a mass of 0.051$\pm$0.006~\Msun\ which
is well below the hydrogen-burning limit.\footnote{
It is well known that the hydrogen-burning limit depends on metallicity \citep{Chabrier97}, with lower metallicity
corresponding to a higher limit. Theoretically and empirically determined limits range from 0.070 to 0.08\,\Msun; below
0.07\,\Msun, brown dwarfs never reach a steady state where they can fuse hydrogen \citep{Fernandes19,Forbes19}.
}
This indicates that BW~Scl might have already evolved through the period minimum. Indeed,
although the formal error of the $M_2$ estimate leaves room for different interpretations, a low temperature of the donor
($\lesssim$1600~K) allows for further restricting its mass and the evolutionary status of the system.
\autoref{Fig:EvolTracks} (right-hand panel) shows a relation between the mass and
temperature of the donor along the `standard' and `best-fitting' evolutionary tracks of CVs from \citet{KniggeCVevol}.
It is apparent from the figure that only a donor star that has passed the period minimum and that has a mass
$\lesssim$0.052~\Msun\ can have so low temperature. In more recent population synthesis studies, it has also been shown
that $T_{\rm eff}$ of the donor after the period bounce is expected to be lower than $\sim$2000\,K \citep{GoliaschNelson},
and that this bounce occurs at $M_{2}$$\approx$0.07\,\Msun\ \citep{Belloni20}. Empirically, using a large set of eclipsing
CVs, \citet{McAllister19} have shown that the bounce occurs at $P_{\rm min}$=76.3$\pm$1.0 min and
$M_{2}=0.063^{+0.005}_{-0.002}$~\Msun. In the $P_{\rm orb}-M_2$ and $T_{\rm eff,2}-M_2$ diagrams (\autoref{Fig:EvolTracks}),
BW~Scl is located below the turning point and, therefore, its donor is a sub-stellar object. Thus, although from the $P_{\rm orb}$
only it is difficult to assess the evolutionary status of BW~Scl because it is located near the theoretical $P_{\rm min}$, all
the above arguments indicate that the system has already passed it and started moving toward longer periods.

A possible period-bouncer status of BW~Scl has been questioned by \citet{Pala22}. They argue that the measured
$T_{\rm eff}$ of its WD is inconsistent with the theoretical predictions for period bouncers and must be $\lesssim$12\,500\,K.
Since $\dot{M}$ is expected to drop as a CV passes beyond $P_{\rm min}$ \citep{KniggeCVevol,GoliaschNelson,Belloni20}, $T_{\rm eff}$
should also fall as it is set by the compressional heating of the accreted material \citep{Townsley04}.
When $T_{\rm eff}$ is measured in quiescence, it provides a constraint on the mean mass-accretion rate
$\langle \dot{M} \rangle$ averaged over the thermal time-scale of the WD envelope ($\sim$$10^{5}$ yr). For BW~Scl with
$T_{\rm eff}$=14250\,K and $M_{\rm wd}$=0.85\,\Msun, equation 2 in \citet{Townsley09} gives
$\langle \dot{M} \rangle$=6.2$\times$10$^{-11}$ \Msun/yr, whereas the most recent population synthesis study by
\citet[][fig. 4]{Belloni20} predicts the range of $\dot{M}$ at $P_{\rm min}$ to be (6.3--8.8)\tim{-11} \Msun/yr.
Thus, both the measured $T_{\rm eff}$ and calculated $\langle \dot{M} \rangle$ are in general agreement with the theory, but,
together with other parameters, testify that BW~Scl is still at the very beginning of its post-$P_{\rm min}$ evolution.

\subsection{Optically thin accretion disc}

Accretion discs in CVs with low-mass accretion rates have outer regions optically thin in continuum \citep{Williams80}.
\citet{Tylenda81} has shown that when lowering $\dot{M}_{\rm acc}$, the outer optically thin region of the disc extends down
to its inner edge (the WD surface), and at $\dot M_{\rm acc}$$\simeq$5\tim{13} g s$^{-1}$ the entire disc becomes optically
thin in continuum. The observed $\dot{M}_{\rm acc}$ in BW~Scl is lower than the above limit implying that most of its accretion
disc, possibly the entire disc is optically thin. As a simple exercise, we can estimate the mean effective (blackbody)
temperature of the disc using the definition of the luminosity as the integral of the total flux over the disc surface:
\begin{equation}
 L_{\rm d} = 2 \pi r^2_{\rm d} \sigma T^4_{\rm eff} \,,
\end{equation}
where $\sigma$ is the Stefan--Boltzmann constant, $r_{\rm d}=r_{\rm d,max}$ is the disc radius, and the factor 2 represents
the radiation from the two sides of the disc. For $L_{\rm d}$$\lesssim$3.2\tim{30} \ergpers\ and $r_{\rm d,max}$=0.338~\Rsun\ we
obtain $T_{\rm eff}$ to be very low, $\sim$2000~K. It is unlikely that so low blackbody temperature represents the true,
kinetic temperature of the disc
material as the latter has to be heated up by e.g. viscosity \citep{Williams80,Tylenda81}. This additionally supports the optically
thin conditions in the disc of BW~Scl. However, the found flat Balmer decrement of the disc (\autoref{Tab:WDparameters})
is indicative of optically thick emission in Balmer lines which are excited rather collisionally than being produced via
photoionization. It suggests that despite the relatively strong irradiating flux from the WD which is able to produce emission
lines from the donor star, yet it is not strong enough to have produced the emission lines from the disc by photoionization alone.

The measured parameters of the Balmer lines can give us some idea of the temperature and density of the line emitting regions.
Comparing the observed Balmer decrements and the EWs of BW~Scl with the model predictions calculated  by \citet{Williams91},
we find that they are roughly consistent with  the disc temperature in the range of 10\,000--15\,000 K and the number density of
hydrogen at the mid-plane of $\log N_0 \approx 12$. However, Williams' radiative transfer models predict much lower EWs than
we observe in BW~Scl, pointing to an even lower density.

There is another, although indirect, evidence for a low-density outer regions of the disc in BW~Scl. It has been shown for several
short-period CVs (WZ~Sge; \citealt{WZ1,WZ2}; HT~Cas; \citealt{NeustroevHT,NeustroevHT2}; EZ~Lyn; \citealt{Amantayeva21}) that
the position of the hotspot in their discs is consistent with the trajectory of the gas stream but is located much closer to the WD
than the disc edge. It is interesting to note that the radial location of the hotspot in those systems is roughly consistent with
the circularization radius $R_{\rm circ}$. In this work we show that this also seems to be correct for BW~Scl, for which
$R_{\rm circ}$=0.39\,$a$ (for $q$=0.060) appears to be very close to $r_{\rm hs}$$\approx$0.35--0.40\,$a$ (Section~\ref{Sec:DopMap}).
Moreover, the Doppler maps of BW~Scl show how the stream propagates through the disc from its very edge to the hotspot. This
structure resembles the so-called hot line, first predicted by \citet{Bisikalo98}. Their numerical simulations show that if the
gas stream is denser than the outer disc then the stream will be able to flow all the way down to the inner disc regions, forming
an extended shock wave along the ballistic trajectory.

\subsection{Optically thick hotspot and double-wave modulations}

The origin of double-wave modulations which are frequently observed in light curves of short-period CVs in quiescence is still
under debate, but it is generally accepted that they are associated with some structures in the accretion disc. In BW~Scl, the
disc contributes $\sim$40 per cent to the total light (in the \textit{TESS} bandpass, Section~\ref{Sec:SED}) which is modulated
with the total amplitude of $\sim$8 per cent (Section~\ref{Sec:TESS}). Thus, these structures are accounted for $\sim$20 per cent
of the accretion disc variability. Two possible interpretations have been suggested to explain this phenomenon.

The first one attributes a double-wave light curve to changing viewing aspect of the optically thick hotspot shining through/above
an optically thin accretion disc \citep{Robinson78,SpruitRutten,WZ1}. This seems to be in accordance with our results. An almost
transparent disc as it is seen in BW~Scl allows the light from the elongated hotspot to escape in all directions, while the variable
aspect of the hotspot modulates the observed flux. We point out that one of the light maxima in BW~Scl occurs at spectroscopic phase
0.86, which is consistent with the phase of maximum of an orbital hump from the stream-disc collision,
typically observed in high-inclination CVs \citep{Warner}.

Another interpretation involves spiral arm structures in the disc which appear due to the 2:1 resonance \citep{Zharikov08}.
In a recent work, \citet{Amantayeva21} have
reported a study of EZ~Lyn, another period-bouncer having many properties similar to BW~Scl. \citeauthor{Amantayeva21}
claim that they were able to successfully reproduce the double-wave light curve of EZ~Lyn, using a model that includes
an accretion disc with a spiral pattern.\footnote{Besides a disc with two spirals, this multicomponent model includes
two extended spots at the outer edge of the disc.}  Nevertheless, besides a broader question of the appearance of the
spirals in the disc, we have other reasons to doubt this interpretation. In order for double-wave modulations to occur, the
spiral structure must be optically thick in the optical light. However, the blackbody temperature of the continuum-forming
region in EZ~Lyn was found to be only 1200--1800~K, while a temperature excess of the spirals (in percent from the disc
temperature) is just $\sim$31 per cent. These results indicate that the spirals should they exist are also optically thin and
thus unable to modulate the observed flux; still, they are treated as optically thick in the mentioned work. Finally, even
assuming that the proposed spirals can modulate the flux, they should manifest themselves in spectral lines, but the very
high-quality Doppler maps of BW~Scl show no sign of them. Instead, the maps display only the bright, optically thick elongated
hotspot (hot line), which cannot help but modulate the signal. Thus, our data do not support the attribution of double-wave
modulations to spiral structures in the disc.

\section{Summary}

We have analysed multi-epoch spectroscopic and photometric observations of the WZ~Sge-type dwarf nova BW Scl, a period-bouncer
candidate. The time-resolved spectroscopic data were obtained in quiescence in 2001, 2002,
2010, 2017 and 2018, before and after the 2011 superoutburst. High-cadence (1 min) X-shooter spectra allowed us to detect
multiple irradiation-induced emission lines from the donor star, permitting the radial velocity variations to be measured with
high accuracy. Using the \MgII\ 4481\,\AA\ and \CaII\,K absorption lines originated in the photosphere of the accreting WD,
we measured its radial velocities and the gravitational redshift, allowing direct measurement of the WD mass.
The most important results of this study can be summarized as follows
\begin{enumerate}
 \item We derived the orbital ephemeris of BW~Scl combining \textit{TESS} and AAVSO photometry and X-shooter spectroscopy.
 \item The WD in BW~Scl has a mass of $M_{\rm wd}$=0.85$\pm$0.04~M$_\odot$.
 \item The measured WD temperature is 14250~K.
 \item The donor is a sub-stellar object with $M_2$=0.051$\pm$0.006~M$_\odot$, well below the hydrogen-burning limit.
 \item Combined NIR spectra, Doppler-corrected into the frame of the donor star, show the \KI\ 12432 and 12522~\AA\ absorption
       lines and hints of a few other species which are expected to be seen in spectra of L and T brown dwarfs. However, their
       presence should be confirmed in further studies.
 \item Using NIR photometric and spectroscopic data, we put a conservative upper limit on the effective temperature of the
       donor to be $T_{\rm eff}\leq$1600~K, corresponding to a brown dwarf of T spectral type.
\item The accretion disc in BW~Scl has a very low luminosity $L_{\rm d}\lesssim$4\tim{30} \ergpers\ which corresponds to
      a very low mass accretion rate of $\dot M_{\rm acc}\lesssim$ 7\tim{-13} \Msun\ yr$^{-1}$. We show that such a disc
      is optically thin in continuum but optically thick in Balmer lines.
\item The outer parts of the disc have a low density allowing the stream to flow down to the inner disc regions. The brightest
      part of the hotspot is located close to the circularization radius of the disc. The hotspot is optically thick and has a
      complex, elongated structure.
\item We suggest that double-wave modulations seen in the light curve of BW~Scl and similar
      objects are produced due changing viewing aspect of the optically thick hotspot shining through an optically thin
      accretion disc.
\item Although the measured donor parameters are consistent with that BW~Scl has already passed the minimum period and started
      moving toward longer periods, the relatively high WD temperature indicates that the system is still at the very beginning
      of its post-$P_{\rm min}$ evolution.
\end{enumerate}

In conclusion, we note that the spectra of BW~Scl demonstrate a wealth of features that allow studying this object in great
detail, permitting us to name this object `a treasure chest' of accreting WDs. Unfortunately, our NIR spectra are not good
enough to investigate the properties of the donor. Accurate measurement of its parameters is of the highest importance as the current
theory is not able to predict physical properties of such a kind of brown dwarf-like objects with
high precision \citep{KniggeCVevol}. BW~Scl is bright enough to allow obtaining deeper NIR spectra than we currently have.

\section*{Acknowledgements}

We would like to thank the anonymous referee for useful comments.
VN acknowledges the financial support from the visitor and mobility program of the Finnish Centre for Astronomy with ESO
(FINCA), funded by the Academy of Finland grant nr 306531.
Based on observations made with ESO Telescopes at the La Silla Paranal Observatory under programme IDs 068.D-0153, 069.D-0391,
086.D-0775, 100.D-0932, and 101.D-0806. This paper includes data collected by the \textit{TESS} mission, which are publicly
available from the Mikulski Archive for Space Telescopes (MAST). Funding for the \textit{TESS} mission is provided by the
NASA's Science Mission Directorate.
This publication makes use of data products from the \textit{Wide-field Infrared Survey Explorer}, which is a joint project
of the University of California, Los Angeles, and the Jet Propulsion Laboratory/California Institute of Technology, funded by
the National Aeronautics and Space Administration. This publication also makes use of data products from NEOWISE, which is a project
of the Jet Propulsion Laboratory/California Institute of Technology, funded by the Planetary Science Division of the National
Aeronautics and Space Administration.
This research made use of StarCAT, hosted by the MAST. STScI is operated by the Association of Universities for Research in
Astronomy, Inc., under NASA contract NAS5-26555.
We thank the {\it Swift} PI, Brad Cenko, for approving the observations, and the {\it Swift}
planning and operations teams for their ongoing support.
We acknowledge with thanks the variable star observations from the {\it AAVSO International Data base} contributed by observers
worldwide and used in this research.

\section*{Data Availability}

The observational data used in this paper are publicly available in the ESO\footnote{\url{http://archive.eso.org}},
NASA/IPAC\footnote{\url{https://irsa.ipac.caltech.edu}}, MAST\footnote{\url{https://mast.stsci.edu}}, and
HEASARC\footnote{\url{https://heasarc.gsfc.nasa.gov/docs/archive.html}} archives.

\bibliographystyle{mnras}
\bibliography{BW_Scl}

\appendix
\section{Supplementary figures}
\label{Sec:Appendix}

\begin{figure*}
\centering
\resizebox{0.7\hsize}{!}{
\includegraphics[angle=0]{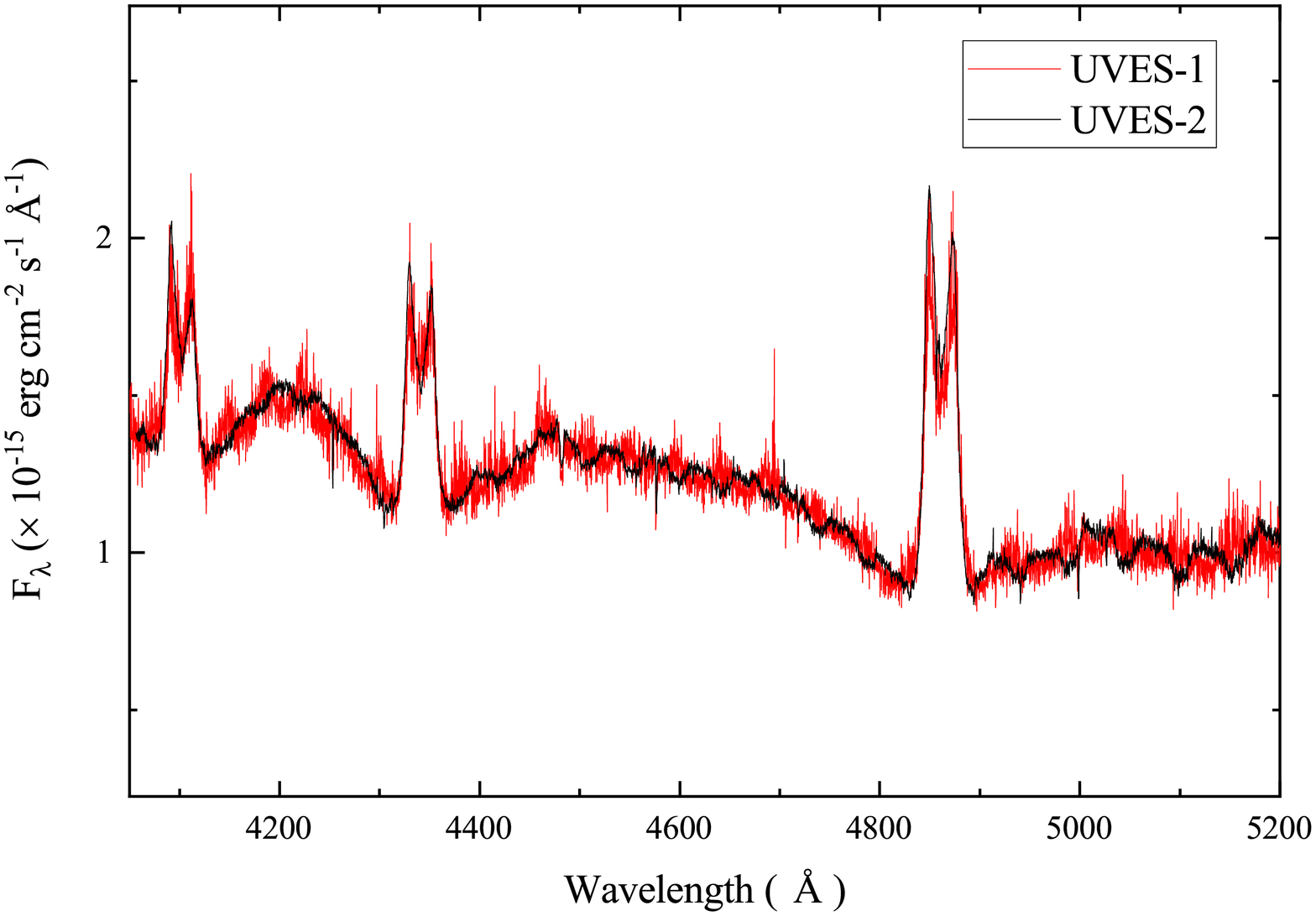}
}
\caption{
The mean UVES spectra clearly exhibit the \'{E}chelle order pattern.
}
\label{Fig:UVES}
\end{figure*}

\begin{figure*}
\centering
\resizebox{\hsize}{!}{
\includegraphics[angle=0]{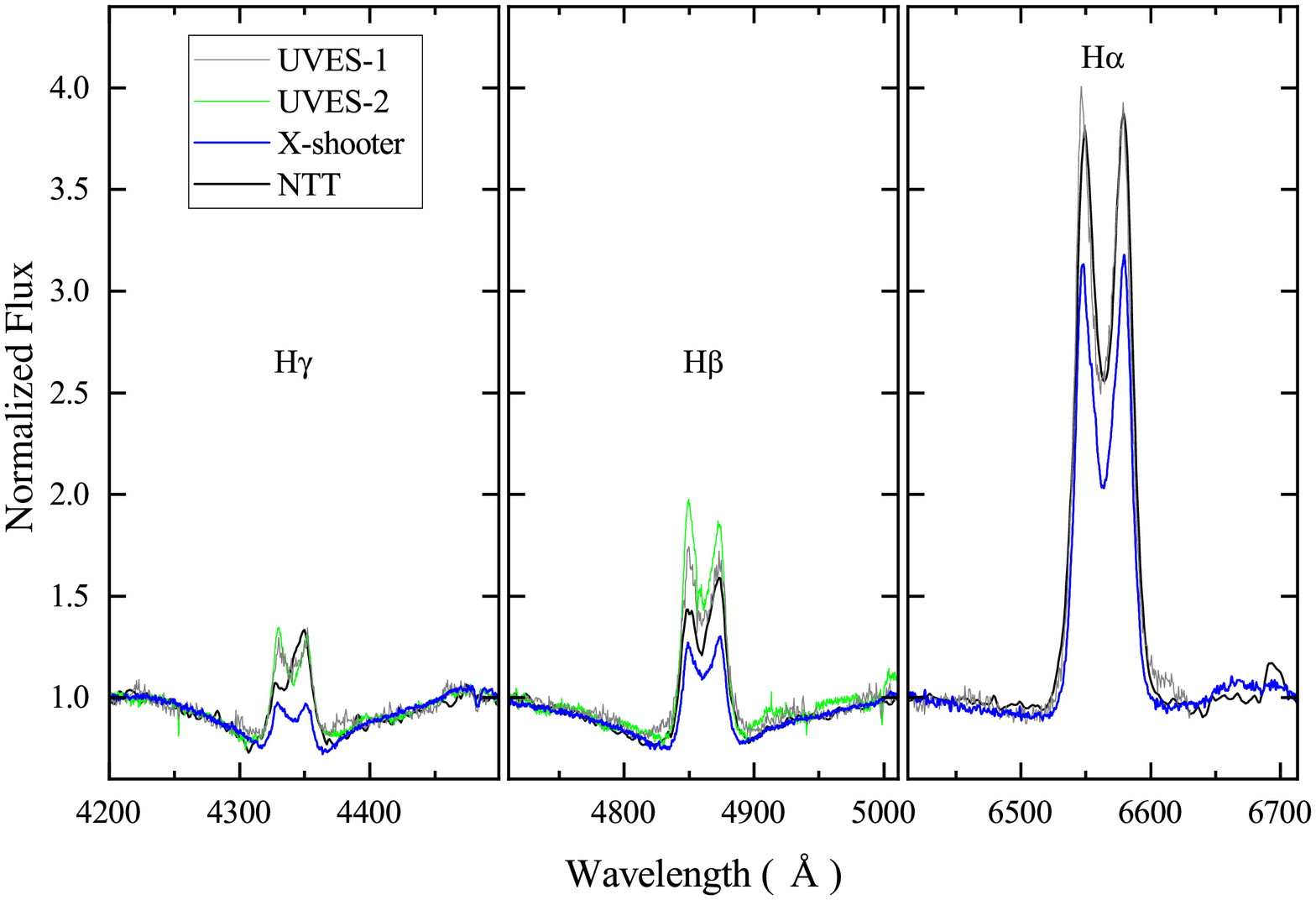}
\includegraphics[angle=0]{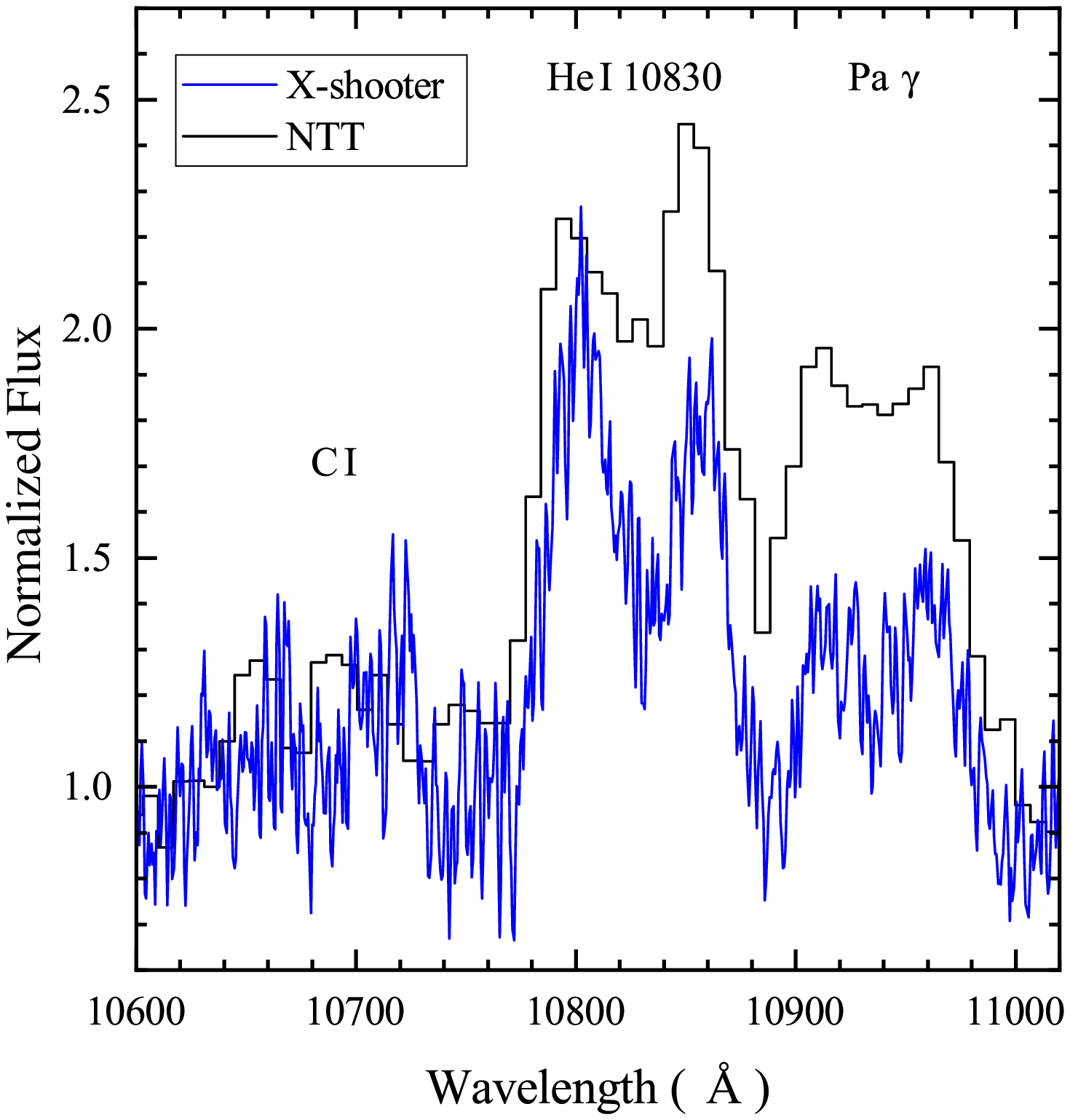}
}
\caption{
\Halpha, \Hbeta, and \Hgamma\ line profiles (left) and the region around the \CI~10693 / \HeI~10830 / \Pagamma\ complex (right)
as seen in different data sets.
}
\label{Fig:SpecOpt-Comp}
\end{figure*}

\begin{figure*}
\resizebox{\hsize}{!}{\includegraphics[angle=0]{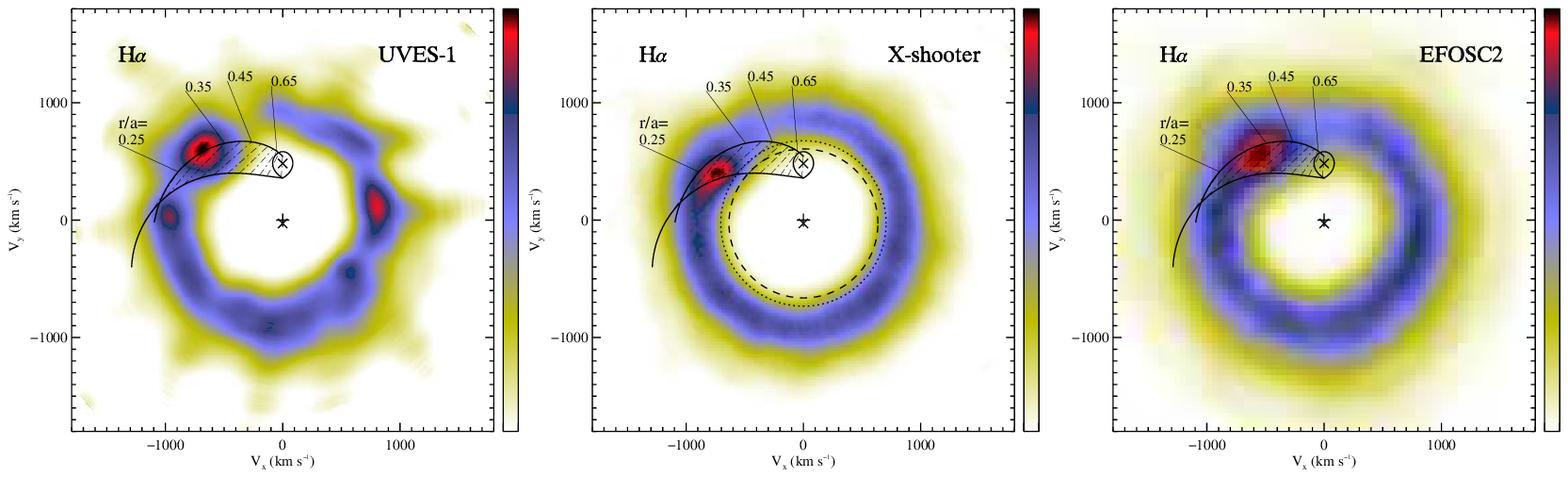}}
    \caption{The Doppler maps of \Halpha\ calculated using different data sets.
             }
    \label{FigApp:DopMapHaAll}
\end{figure*}


\begin{figure*}
\resizebox{\hsize}{!}{\includegraphics[angle=0]{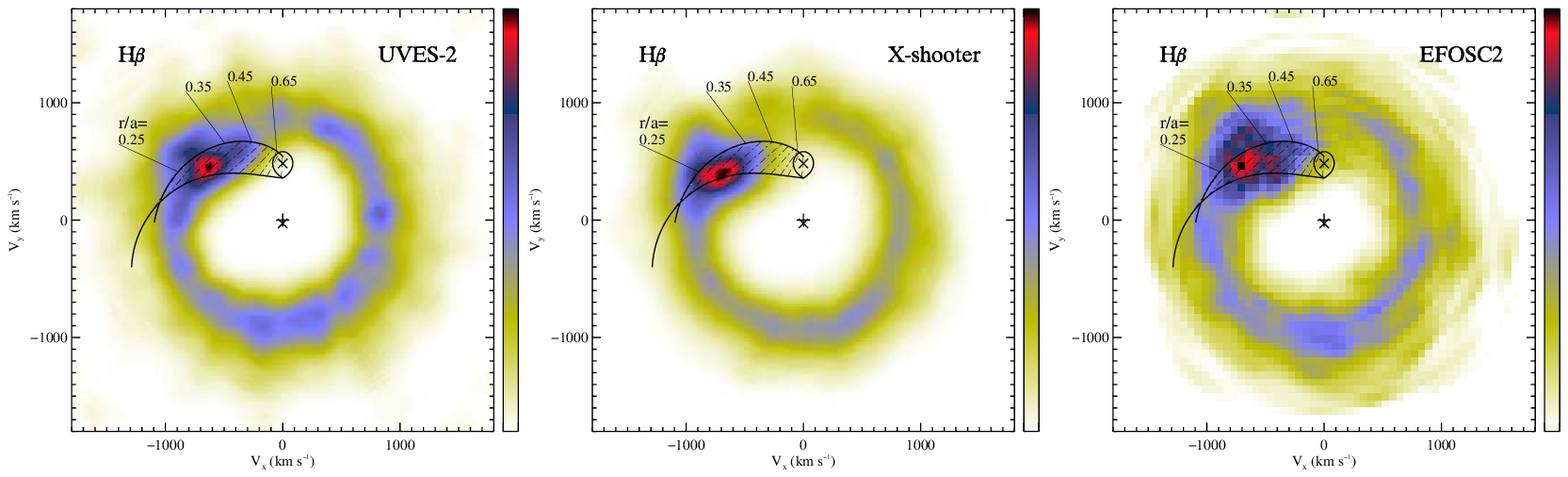}}
    \caption{The Doppler maps of \Hbeta\ calculated using different data sets.
             }
    \label{FigApp:DopMapHbAll}
\end{figure*}

\begin{figure*}
\resizebox{\hsize}{!}{\includegraphics[angle=0]{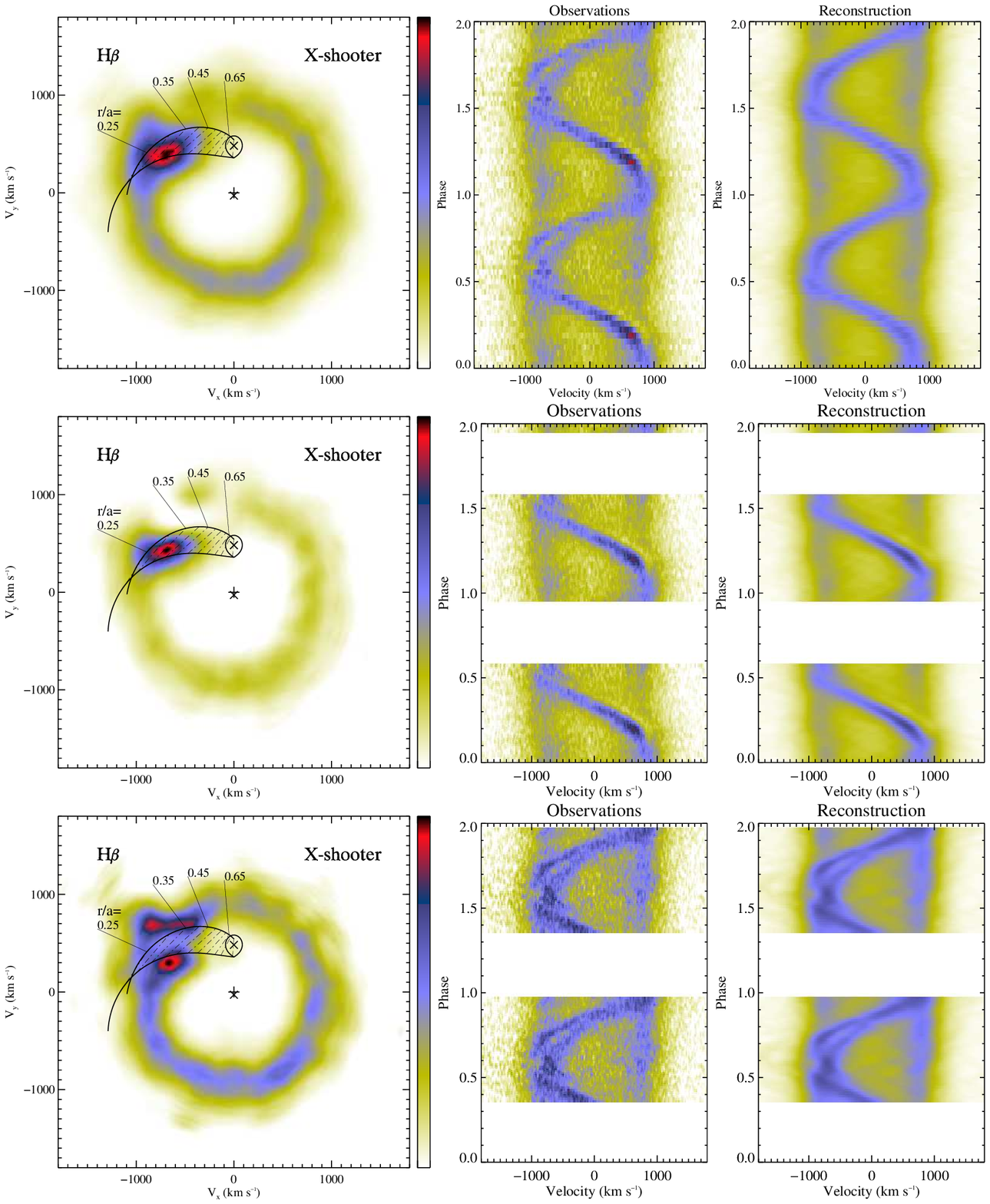}}
    \caption{Same as \autoref{Fig:DopMapHa}, for \Hbeta. The middle and lower maps were calculated using 60 per cent of the spectra
    centred on phases 0.25 and 0.65, respectively.
             }
    \label{Fig:DopMapHb}
\end{figure*}

\begin{figure*}
    \resizebox{\hsize}{!}{\includegraphics[angle=0]{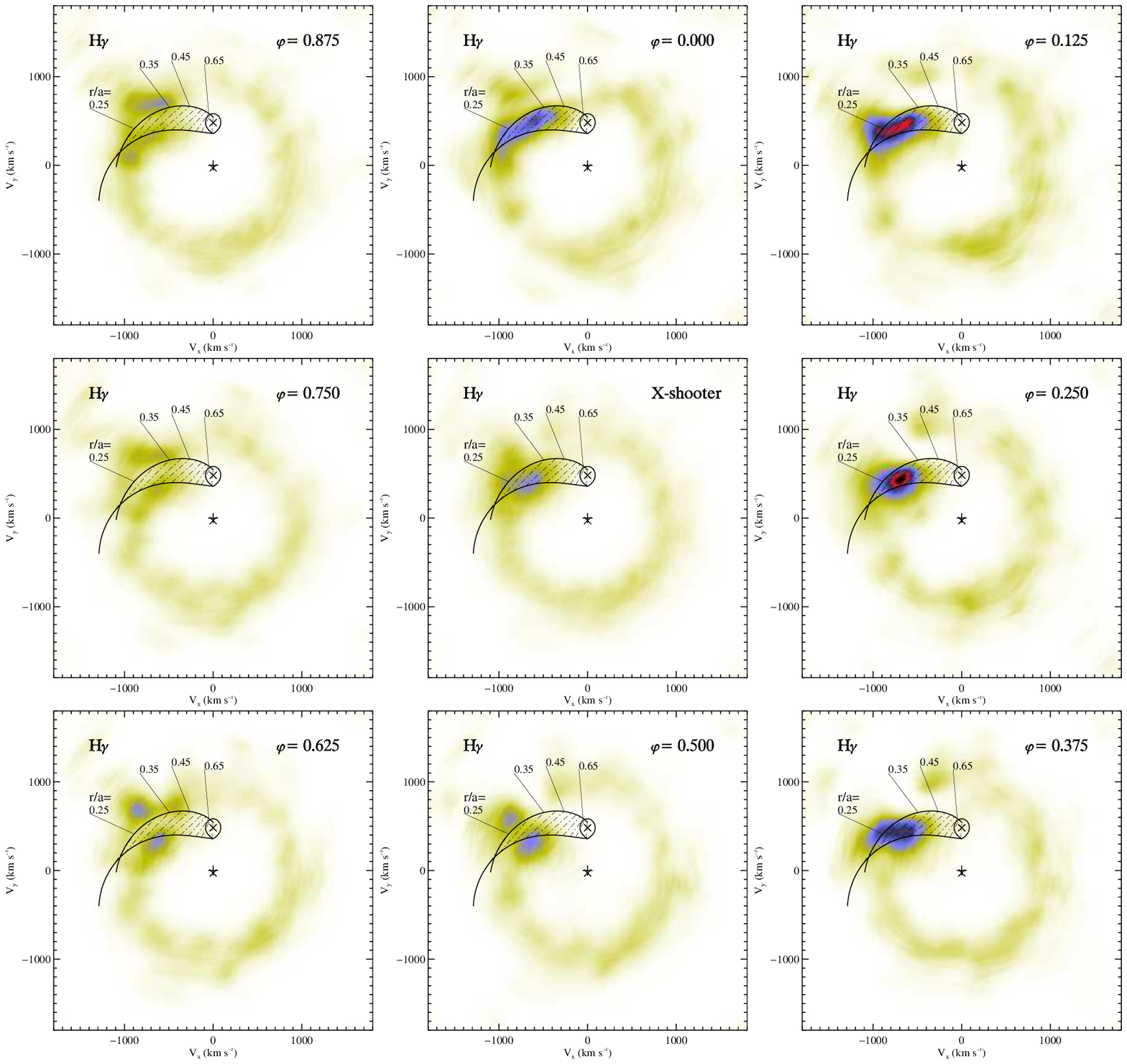}}
    \caption{Same as \autoref{Fig:DopMapDynHS}, but the full velocity range maps were scaled according to the average disc brightness.
             }
    \label{Fig:DopMapDynDisc}
\end{figure*}

\bsp	
\label{lastpage}
\end{document}